\documentclass[11pt]{article}
\usepackage{amssymb,amsmath,amsfonts}
\usepackage{graphicx}
\usepackage{graphics}
\usepackage{eepic,epsfig}

\@addtoreset{equation}{section}

\makeatother

\begin{document}

\title{Electromagnetic Casimir densities for a cylindrical \\
shell on de Sitter space}
\author{A. A. Saharian$^{1}$,\thinspace\ V. F. Manukyan$^{2}$,\thinspace\ N.
A. Saharyan$^{1}$ \\
\\
\textit{$^1$Department of Physics, Yerevan State University,}\\
\textit{1 Alex Manoogian Street, 0025 Yerevan, Armenia}\vspace{0.3cm}\\
\textit{$^{2}$Department of Physics and Mathematics, Gyumri State
Pedagogical Institute,}\\
\textit{4 Paruyr Sevak Street, 3126 Gyumri, Armenia}}
\maketitle

\begin{abstract}
Complete set of cylindrical modes is constructed for the electromagnetic
field inside and outside a cylindrical shell in the background of $(D+1)$%
-dimensional dS spacetime. On the shell, the field obeys the generalized
perfect conductor boundary condition. For the Bunch-Davies vacuum state, we
evaluate the expectation values (VEVs) of the electric field squared and of
the energy-momentum tensor. The shell-induced contributions are explicitly
extracted. In this way, for points away from the shell, the renormalization
is reduced to the one for the VEVs in the boundary-free dS bulk. As a
special case, the VEVs are obtained for a cylindrical shell in the $(D+1)$%
-dimensional Minkowski bulk. We show that the shell-induced contribution in
the electric field squared is positive for both the interior and exterior
regions. The corresponding Casimir-Polder forces are directed toward the
shell. The vacuum energy-momentum tensor, in addition to the diagonal
components, has a nonzero off-diagonal component corresponding to the energy
flux along the direction normal to the shell. This flux is directed from the
shell in both the exterior and interior regions. For points near the shell,
the leading terms in the asymptotic expansions for the electric field
squared and diagonal components of the energy-momentum tensor are obtained
from the corresponding expressions in the Minkowski bulk replacing the
distance from the shell by the proper distance in the dS bulk. The influence
of the gravitational field on the local characteristics of the vacuum is
essential at distances from the shell larger than the dS curvature radius.
The results are extended for confining boundary conditions of flux tube
models in QCD.
\end{abstract}

\bigskip


\bigskip

\section{Introduction}

The Casimir effect (for reviews see \cite{Most97}) is among the most
interesting boundary-induced effects in quantum field theory. Nearly seventy
years after its discovery, the effect continues to be area of active
theoretical and experimental research today. This is due to of fundamental
and practical importance of the Casimir effect in various fields of modern
physics and technology. The imposition of boundary conditions on a quantum
field shifts the expectation values of physical observables. The particular
features of the effect depend on the spin of the field, on the geometries of
boundaries and background spacetime, and on the specific boundary
conditions. An interesting topic in the investigations of the Casimir effect
is the dependence of the physical characteristics, such as the vacuum energy
and forces acting on the boundaries, on the dimension of the background
spacetime. In particular, this is motivated by the role of the Casimir
effect in higher-dimensional models. The dependence of the Casimir energy on
the size of the extra dimensions provides a mechanism for their
stabilization. The latter is an important issue in both the
Kaluza-Klein-type and braneworld models. In addition, the Casimir energy
related to extra dimensions can serve as a model of dark energy driving the
accelerating expansion of the Universe at recent epoch.

In the present paper we investigate combined effects of boundaries and
background geometry on the local characteristics of the electromagnetic
vacuum in an arbitrary number of spatial dimensions. As a background
spacetime we consider $(D+1)$-dimensional de Sitter (dS) spacetime. This
choice is motivated by several reasons. The dS spacetime has maximal number
of symmetries and because of this a large number of physical problems can be
solved exactly on its background. A better understanding of the influence of
classical gravitational field on quantum matter in dS spacetime serves as a
way to deal with less symmetric geometries. The quantum effects in dS bulk
play an important role in inflationary scenarios (see \cite{Lind94} and \cite%
{Mart13} for recent reviews). Quantum fluctuations in the inflationary
universe with dS geometry produce observable effects in the form of small
density perturbations in the post-inflationary stage. These perturbations
have nearly scale-invariant spectrum and form the base of the currently most
popular mechanism for the generation of the seeds for galaxy formation. The
corresponding predictions are in good agreement with the recent
observational data on the temperature anisotropies of the cosmic microwave
background radiation. The importance of dS spacetime as a gravitational
background further increased after the discovery of the accelerating
expansion of the Universe at recent epoch \cite{Ries98,Frie08}. Within the
framework of general relativity, among the most popular cosmological models
is the one with a positive cosmological constant as a driving source behind
the accelerating expansion. In this model the dS spacetime is the future
attractor for the geometry of our universe.

An important ingredient in the problems of the Casimir effect is the
geometry of boundaries. Here we consider a cylindrical shell in background
of dS spacetime, described by inflationary coordinates. On the shell the
electromagnetic field obeys the generalized perfect conductor boundary
condition. Along with the boundaries of planar and spherical symmetry, the
cylindrical boundaries are among the most popular geometries in the
investigations of the Casimir effect (see \cite{Most97} and references
therein). The patch of dS spacetime covered by inflationary coordinates is
conformally flat and the corresponding results for the Casimir effect in the
case of conformally invariant fields are obtained from the expressions for
the Minkowski bulk by using the standard conformal relation between two
problems (see, for instance, \cite{Birr82}). In particular, this is the case
for the electromagnetic field in 4-dimensional spacetime. The case of
conformally coupled massless scalar field obeying Dirichlet boundary
condition on a spherical shell in 4-dimensional dS spacetime has been
considered in \cite{Seta01}.

For conformally non-invariant fields the investigation of the Casimir effect
in dS spacetime requires a separate consideration. In this case, at
distances from the boundaries larger than the curvature radius of the
background spacetime, the influence of the gravitational field may lead to a
behavior of the local characteristics of the vacuum essentially different
from that in the Minkowski bulk. The scalar Casimir effect for a massive
field with general curvature coupling and with Robin boundary conditions on
planar boundaries in the background of $(D+1)$-dimensional dS spacetime has
been investigated in \cite{Saha09}. The corresponding problem with a
spherical boundary was discussed in \cite{Milt12}. An important feature for
conformally non-invariant fields is that the vacuum energy-momentum tensor,
in addition to the diagonal components, acquires an off-diagonal component
corresponding to the energy flux along the direction normal to the boundary.
The Casimir densities induced by a spherical dS bubble in the Minkowski bulk
has been studied in \cite{Bell14}. For a scalar field, the geometry with a
cylindrical boundary with Robin boundary condition was discussed in \cite%
{Saha15}. Depending on the mass of the field, at distances from the
boundaries larger than the dS curvature radius, two different regimes are
realized with monotonic or oscillatory decay of the vacuum expectation
values. The electromagnetic Casimir effect for planar boundaries with
generalized perfect conductor boundary conditions in dS bulk, having an
arbitrary number of spatial dimensions, has recently been studied in \cite%
{Saha14} (for the propagators of vector fields on dS background in the
absence of boundaries see \cite{Alle86}). The background geometry of
Friedmann-Robertson-Walker cosmologies with power-law scale factors was
discussed in \cite{Bell13}.

The organization of the paper is the following. In the next section we
consider the complete set of cylindrical modes for the electromagnetic field
in dS spacetime, obeying the generalized perfect conductor boundary
condition on a cylindrical shell. By using these modes, in section \ref%
{sec:E2}, the vacuum expectation value (VEV) of the electric field squared
inside the shell is evaluated. Various asymptotic regions of the parameters
are discussed. The VEV of the energy-momentum tensor is studied in section %
\ref{sec:EMT}. The VEVs of the electric field squared and of the
energy-momentum tensor in the exterior region are investigated in section %
\ref{sec:Ext}. Section \ref{sec:Conc}\ summarizes the main results of the
paper. In appendix \ref{sec:App1} the cylindrical modes in $(D+1)$%
-dimensional Minkowski spacetime are discussed. The integrals, appearing in
the expressions for the VEVs for the special case $D=4$, are evaluated in
appendix \ref{sec:App2}. In the numerical evaluations of the VEVs we have
considered this special case.

\section{Electromagnetic field modes in dS spacetime}

\label{sec:Modes}

We consider a quantum electromagnetic field in background of a $(D+1)$%
-dimensional dS spacetime, in the presence of a perfectly conducting
cylindrical shell having the radius $a$. In accordance with the problem
symmetry, we write the dS line element in cylindrical coordinates $\left(
r,\phi ,\mathbf{z}\right) $:
\begin{equation}
ds^{2}=dt^{2}-e^{2t/\alpha }\left[ dr^{2}+r^{2}d\phi ^{2}+\left( d\mathbf{z}%
\right) ^{2}\right] ,  \label{ds2}
\end{equation}%
where $\mathbf{z=}\left( z^{3},...,z^{D}\right) $. The parameter $\alpha $
is expressed in terms of the cosmological constant $\Lambda $ through the
relation $\alpha ^{2}=D(D-1)/(2\Lambda )$. Below, in addition to the
synchronous time coordinate $t$ we will use the conformal time $\tau $,
defined as $\tau =-\alpha e^{-t/\alpha },\ -\infty <\tau <0$. In terms of
this coordinate the metric tensor takes a conformally flat form $g_{\mu \nu
}=(\alpha /\tau )^{2}g_{(\mathrm{M})\mu \nu }$, where $g_{(\mathrm{M})\mu
\nu }=\mathrm{diag}(1,-1,-r^{2},-1,\ldots ,-1)$ is the Minkowskian metric
tensor.

We are interested in the changes of the VEVs for the electromagnetic field
induced by a cylindrical boundary $r=a$ in the background of the geometry
given by (\ref{ds2}). In the canonical quantization procedure a complete
orthonormal set of solutions to the classical field equations is required
and, in this section, we present this set for the geometry at hand. For a
free electromagnetic field the Maxwell equations have the form
\begin{equation}
\frac{1}{\sqrt{|g|}}\partial _{\nu }\left( \sqrt{|g|}F^{\mu \nu }\right) =0,
\label{Meq}
\end{equation}%
where $F_{\mu \nu }$ is the electromagnetic field tensor, $F_{\mu \nu
}=\partial _{\mu }A_{\nu }-\partial _{\nu }A_{\mu }$. We assume that on the
surface $r=a$ the field obeys the boundary condition
\begin{equation}
n^{\nu _{1}}\,^{\ast }F_{\nu _{1}\cdots \nu _{D-1}}=0,  \label{BC1}
\end{equation}%
where $n^{v}$ is the normal to the boundary, $^{\ast }F_{\nu _{1}\cdots \nu
_{D-1}}$ is the dual of the field tensor $F_{\mu \nu }$. For $D=3$ the
condition (\ref{BC1}) reduces to the boundary condition on the surface of a
perfect conductor. We want to find the complete set of solutions to the
equation (\ref{Meq}) in the coordinates $(\tau ,r,\phi ,\mathbf{z})$.

In the Coulomb gauge one has $A_{0}=0$, $\partial _{l}(\sqrt{|g|}A^{l})=0$, $%
l=1,...,D$. For the geometry under consideration the latter equation is
reduced to%
\begin{equation}
\sum_{l=1}^{D}\partial _{l}(rA^{l})=0,  \label{Gcond}
\end{equation}%
which is the same as that in the Minkowski spacetime. If we present the
solution in the factorized form, $A_{\mu }(\tau ,x^{l})=T(\tau )S_{\mu
}(x^{l})$, then it can be shown that the parts of the mode functions
corresponding to $S_{\mu }(x^{l})$ are found in a way similar to that for
the $(D+1)$-dimensional Minkowski bulk. The corresponding modes are
presented in appendix \ref{sec:App1}. From the field equations (\ref{Meq})
for the function $T(\tau )$ one gets $T(\tau )=\eta ^{D/2-1}Z_{D/2-1}(\omega
\eta )$, where $\eta =|\tau |$ and $Z_{\nu }(x)$ is a cylinder function of
the order $\nu $. It can be taken as a linear combination of the Hankel
functions $H_{\nu }^{(1,2)}(x)$. The relative coefficient in the linear
combination depends on the vacuum state under consideration. We assume that
the field is prepared in the Bunch-Davies vacuum \cite{Bunc78} for which $%
Z_{\nu }(x)=H_{\nu }^{(1)}(x)$. Among a one-parameter family of maximally
symmetric vacuum states in dS spacetime the Bunch-Davies vacuum is the only
one with the Hadamard singularity structure.

As a result, by taking into account the expressions for the Minkwoskian
modes (\ref{A1M}) and (\ref{AmuM}), for the modes in dS spacetime realizing
the Bunch-Davies vacuum state one finds the expressions%
\begin{equation}
A_{(\beta )\mu }=c_{\beta }\eta ^{D/2-1}H_{D/2-1}^{(1)}(\omega \eta
)(0,im/r,-r\partial _{r},0,\ldots ,0)C_{m}(\gamma r)e^{im\phi +i\mathbf{k}%
\cdot \mathbf{z}},  \label{A1}
\end{equation}%
for the polarization $\sigma =1$, and
\begin{equation}
A_{(\beta )\mu }=\omega c_{\beta }\eta ^{D/2-1}H_{D/2-1}^{(1)}(\omega \eta
)\left( 0,\epsilon _{\sigma l}+i\omega ^{-2}\mathbf{k}\cdot \mathbf{\epsilon
}_{\sigma }\partial _{l}\right) C_{m}(\gamma r)e^{im\phi +i\mathbf{k}\cdot
\mathbf{z}},  \label{A2}
\end{equation}%
for the remaining polarizations $\sigma =2,\ldots ,D-1$, with $l=1,\ldots ,D$%
. Here, $m=0,\pm 1,\pm 2,\ldots $, $C_{m}(x)$ is a cylinder function, $%
\mathbf{k}\cdot \mathbf{z}=\sum_{l=3}^{D}k_{l}z^{l}$, $\mathbf{k}\cdot
\mathbf{\epsilon }_{\sigma }=\sum_{l=3}^{D}k_{l}\epsilon _{\sigma l}$, and%
\begin{equation}
\omega =\sqrt{\gamma ^{2}+k^{2}},  \label{om}
\end{equation}%
with $k^{2}=\sum_{l=3}^{D}k_{l}^{2}$. For the polarization vectors $\epsilon
_{\sigma l}$ one has $\epsilon _{\sigma 1}=\epsilon _{\sigma 2}=0$ and the
relations (\ref{rel1M}), (\ref{rel2M}). The set of quantum numbers
specifying the modes is reduced to $\beta =(\gamma ,m,\mathbf{k},\sigma )$.
It can be easily checked that the modes (\ref{A1}) and (\ref{A2}) obey the
gauge condition (\ref{Gcond}).

The eigenvalues for the quantum number $\gamma $ are determined by the
boundary condition (\ref{BC1}). First we consider the region inside the
cylindrical shell, $r<a$. From the regularity condition at $r=0$ it follows
that%
\begin{equation}
C_{m}(\gamma r)=J_{m}(\gamma r),  \label{RadIn}
\end{equation}%
with $J_{m}(x)$ being the Bessel function. For the mode $\sigma =1$ the
allowed values for $\gamma $ are roots of the equation%
\begin{equation}
J_{m}^{\prime }(\gamma a)=0,  \label{BCa}
\end{equation}%
where the prime means the derivative with respect to the argument of the
function. For the modes $\sigma =2,\ldots ,D-1$, the radial derivative
enters in the expression for the component $A_{1}$ only. For these modes the
boundary condition is reduced to%
\begin{equation}
J_{m}(\gamma a)=0.  \label{BCb}
\end{equation}%
In what follows we will denote the eigenmodes by $\gamma a=\gamma
_{m,n}^{(\lambda )}$,\ $n=1,2,...$, where $\lambda =1$ for $\sigma =1$ and $%
\lambda =0$ for $\sigma =2,3,\ldots ,D-1$. Hence, one has $J_{m}^{(\lambda
)}(\gamma _{m,n}^{(\lambda )})=0$, with $f^{(0)}(x)=f(x)$ and $%
f^{(1)}(x)=f^{\prime }(x)$.

The normalization coefficients $c_{\beta }$ in (\ref{A1}) and (\ref{A2}) are
determined from the orthonormalization condition for the vector potential:
\begin{equation}
\int d^{D}x\sqrt{|g|}[A_{(\beta ^{\prime })\nu }^{\ast }(x)\nabla
^{0}A_{\beta }^{\nu }(x)-\nabla ^{0}A_{(\beta ^{\prime })\nu }^{\ast
}(x)A_{(\beta )}^{\nu }(x)]=4i\pi \delta _{\beta \beta ^{\prime }},
\label{NormCond}
\end{equation}%
where the integration over the radial coordinate goes over the region inside
the cylinder and $\delta _{\beta \beta ^{\prime }}$ is understood as the
Kronecker symbol for the discrete components of the collective index $\beta $
and the Dirac delta function for the continuous ones. By using the relation (%
\ref{rel1M}) it can be seen that the modes (\ref{A2}) are orthogonal. From (%
\ref{NormCond}) one finds
\begin{equation}
|c_{\beta }|^{2}=\frac{T_{m}(\gamma _{m,n}^{(\lambda )})}{2(2\pi \alpha
)^{D-3}\gamma _{m,n}^{(\lambda )}},  \label{Csig}
\end{equation}%
for $\sigma =1,\ldots ,D-1$, where we have introduced the notation%
\begin{equation}
T_{m}(x)=x\left[ x^{2}J_{m}^{\prime 2}(x)+\left( x^{2}-m^{2}\right)
J_{m}^{2}(x)\right] ^{-1}.  \label{Tnu}
\end{equation}

Having the complete set of mode functions (\ref{A1}) and (\ref{A2}) we can
evaluate the VEV of any physical quantity $F\{A_{\mu }(x),A_{\nu }(x)\}$
bilinear in the field. By expanding the operator of the vector potential in
terms of the modes (\ref{A1}) and (\ref{A2}) and using the commutation
relations for the annihilation and creation operators, the following
mode-sum formula is obtained%
\begin{equation}
\langle 0|F\{A_{\mu }(x),A_{\nu }(x)\}|0\rangle =\sum_{\beta }F\{A_{(\beta
)\mu }(x),A_{(\beta )\nu }^{\ast }(x)\},  \label{VEV}
\end{equation}%
where $|0\rangle $ stands for the vacuum state, $\sum_{\beta }$ includes the
summation over the discrete quantum numbers and the integration over the
continuous ones. The expression in the right-hand side of (\ref{VEV}) is
divergent and requires a regularization with the subsequent renormalization.
The regularization can be done by introducing a cutoff function or by the
point splitting. The consideration below does not depend on the specific
regularization scheme and we will not specify it.

\section{VEV of the electric field squared}

\label{sec:E2}

As a local characteristic of the vacuum state we consider the VEV\ of the
squared electric field. This VEV is obtained by making use of the mode-sum
formula
\begin{equation}
\langle 0|E^{2}(x)|0\rangle \equiv \langle E^{2}(x)\rangle
=-g^{00}g^{il}\sum_{\beta }\partial _{0}A_{(\beta )i}(x)\partial
_{0}A_{(\beta )l}^{\ast }(x).  \label{E2mode}
\end{equation}%
Note that the VEV of the electric field squared determines the
Casimir-Polder potential between the shell and a polarizable particle with a
frequency-independent polarizability. Substituting the eigenfunctions (\ref%
{A1}) and (\ref{A2}), after the summation over $\sigma $ with the help of (%
\ref{rel2M}), for the VEV inside the shell we find
\begin{eqnarray}
\langle E^{2}\rangle  &=&\frac{2^{6}A_{D}\eta ^{D+2}}{\alpha ^{D+1}a^{4}}\ %
\sideset{}{'}{\sum}_{m=0}^{\infty }\int_{0}^{\infty
}dkk^{D-3}\sum\limits_{\lambda =0,1}\sum_{n=1}^{\infty }T_{m}(\gamma
_{m,n}^{(\lambda )})  \notag \\
&&\times \gamma _{m,n}^{(\lambda )3}F_{m}^{(\lambda )}[k,J_{m}(\gamma
_{m,n}^{(\lambda )}r/a)]L_{D/2-2}(\omega _{m,n}^{(\lambda )}\eta ),
\label{E2}
\end{eqnarray}%
where $\omega _{m,n}^{(\lambda )}=\sqrt{\gamma _{m,n}^{(\lambda
)}/a^{2}+k^{2}}$, the prime on the sign of the sum means that the term $m=0$
should be taken with the coefficient $1/2$,%
\begin{equation}
A_{D}=\frac{1}{(4\pi )^{D/2}\Gamma (D/2-1)},  \label{AD}
\end{equation}%
and we have used the notation%
\begin{equation}
L_{\nu }(x)=K_{\nu }(xe^{-i\pi /2})K_{\nu }(xe^{i\pi /2}).  \label{Lnu}
\end{equation}%
Here, instead of the Hankel functions we have introduced the Macdonald
function $K_{\nu }(x)$. The function $F_{m}^{(\lambda )}[k,f(x)]$ is defined
by the relations%
\begin{equation}
F_{m}^{(\lambda )}\left[ k,f(x)\right] =\left\{
\begin{array}{cc}
(k^{2}r^{2}/x^{2})\left[ f^{\prime 2}(x)+m^{2}f^{2}(x)/x^{2}\right] +\left[
\left( D-3\right) \left( 1+k^{2}r^{2}/x^{2}\right) +1\right] f^{2}(x), &
\lambda =0, \\
(1+k^{2}r^{2}/x^{2})\left[ f^{\prime 2}(x)+m^{2}f^{2}(x)/x^{2}\right] , &
\lambda =1.%
\end{array}%
\right.   \label{Gm}
\end{equation}

The eigenvalues $\gamma _{m,n}^{(\lambda )}$ are given implicitly and the
representation (\ref{E2}) is not convenient for the further investigation of
the VEV. For the further evaluation of the mode-sum in (\ref{E2}), we apply
to the series over $n$ the generalized Abel-Plana summation formula \cite%
{SahaBook}
\begin{eqnarray}
&&\sum_{n=1}^{\infty }T_{m}(\gamma _{m,n}^{(\lambda )})f(\gamma
_{m,n}^{(\lambda )})=\frac{1}{2}\int_{0}^{\infty }dxf\left( x\right) +\frac{%
\pi }{4}{\mathrm{Res}}_{z=0}f\left( z\right) \frac{Y_{m}^{\left( \lambda
\right) }\left( z\right) }{J_{m}^{\left( \lambda \right) }\left( z\right) }
\notag \\
&&\qquad -\frac{1}{2\pi }\int_{0}^{\infty }dz\,\dfrac{K_{m}^{\left( \lambda
\right) }\left( z\right) }{I_{m}^{\left( \lambda \right) }\left( z\right) }%
\left[ e^{-m\pi i}f(ze^{i\pi /2})+e^{m\pi i}f(ze^{-i\pi /2})\right] ,
\label{SumForm}
\end{eqnarray}%
where $Y_{m}\left( z\right) $ is the Neumann function and $I_{m}\left(
x\right) $ is the modified Bessel function of the first kind. In (\ref%
{SumForm}) it is assumed that the function $f\left( z\right) $ is analytic
in the right half plane of the complex variable $z$. For the series in (\ref%
{E2}), the corresponding function $f\left( z\right) $ is given by%
\begin{equation}
f\left( x\right) =x^{3}L_{D/2-2}(\eta \sqrt{x^{2}/a^{2}+k^{2}})F_{m}^{\left(
\lambda \right) }\left[ k,J_{m}\left( xr/a\right) \right] .  \label{fz}
\end{equation}%
Note that this function has branch points $x=\pm ika$ on the imaginary axis.
For the function (\ref{fz}) one has the relation%
\begin{equation}
\sum_{j=-1,1}e^{jm\pi i}f(ze^{-ji\pi /2})=ie^{-m\pi i}z^{3}F_{m}^{\left(
\lambda \right) }\left[ k,J_{m}\left( ze^{i\pi /2}r/a\right) \right]
\sum_{j=-1,1}jL_{D/2-2}(\eta \sqrt{z^{2}e^{-ji\pi }/a^{2}+k^{2}}),
\label{Sumf}
\end{equation}%
with%
\begin{equation}
\sum_{j=-1,1}jL_{\nu }(\eta \sqrt{z^{2}e^{-ji\pi }/a^{2}+k^{2}})=i\pi
\left\{
\begin{array}{cc}
0, & z<ka \\
f_{\nu }(\eta \sqrt{z^{2}/a^{2}-k^{2}}), & z>ka%
\end{array}%
\right. .  \label{SumL}
\end{equation}%
Here and in what follows we use the notation%
\begin{equation}
f_{\nu }(x)=K_{\nu }\left( x\right) \left[ I_{-\nu }\left( x\right) +I_{\nu
}\left( x\right) \right] .  \label{gnu}
\end{equation}

After the application of the summation formula (\ref{SumForm}), the VEV of
the electric field squared is decomposed as
\begin{equation}
\langle E^{2}\rangle =\langle E^{2}\rangle _{\mathrm{dS}}+\langle
E^{2}\rangle _{\mathrm{b}}.  \label{E2Dec}
\end{equation}%
Here, the first term in the right-hand side comes from the first integral in
(\ref{SumForm}). It does not depend on the shell radius $a$ and corresponds
to the VEV in dS spacetime in the absence of boundaries:
\begin{equation}
\langle E^{2}\rangle _{\mathrm{dS}}=\frac{2^{5}A_{D}\eta ^{D+2}}{\alpha
^{D+1}}\ \sideset{}{'}{\sum}_{m=0}^{\infty }\int_{0}^{\infty
}dk\,k^{D-3}\int_{0}^{\infty }dx\,x^{3}F_{m}[k,J_{m}\left( xr\right)
]L_{D/2-2}(\eta \sqrt{k^{2}+x^{2}}),  \label{E20}
\end{equation}%
where%
\begin{eqnarray}
F_{m}\left[ k,f(x)\right] &=&\left( 1+2\frac{k^{2}r^{2}}{x^{2}}\right) \left[
f^{\prime 2}(x)+\frac{m^{2}}{x^{2}}f^{2}(x)\right]  \notag \\
&&+\left[ \left( D-3\right) \left( 1+\frac{k^{2}r^{2}}{x^{2}}\right) +1%
\right] f^{2}(x).  \label{G0}
\end{eqnarray}%
The part of the VEV $\left\langle E^{2}\right\rangle _{\mathrm{b}}$ is the
contribution of the last integral in (\ref{SumForm}). This part is induced
by the presence of the cylindrical boundary and is given by the expression
\begin{eqnarray}
\left\langle E^{2}\right\rangle _{\mathrm{b}} &=&\frac{2^{5}A_{D}}{\alpha
^{D+1}}\sideset{}{'}{\sum}_{m=0}^{\infty }\sum\limits_{\lambda
=0,1}\int_{0}^{\infty }dx\,x^{D+1}\dfrac{K_{m}^{\left( \lambda \right)
}\left( xa/\eta \right) }{I_{m}^{\left( \lambda \right) }\left( xa/\eta
\right) }  \notag \\
&&\times \int_{0}^{1}ds\,s\left( 1-s^{2}\right) ^{D/2-2}G_{m}^{\left(
\lambda \right) }\left[ s,I_{m}\left( xr/\eta \right) \right]
f_{D/2-2}\left( xs\right) ,  \label{E2b}
\end{eqnarray}%
with the notation%
\begin{equation}
G_{m}^{(\lambda )}\left[ s,f(x)\right] =\left\{
\begin{array}{cc}
\left( 1-s^{2}\right) \left[ f^{\prime 2}(x)+m^{2}f^{2}(x)/x^{2}\right] +%
\left[ s^{2}\left( D-3\right) +1\right] f^{2}(x), & \lambda =0, \\
-s^{2}\left[ f^{\prime 2}(x)+m^{2}f^{2}(x)/x^{2}\right] , & \lambda =1.%
\end{array}%
\right.  \label{Gb}
\end{equation}%
In deriving (\ref{E2b}), after using (\ref{Sumf}) and (\ref{SumL}), we have
introduced a new integration variable $u$ in accordance with $z=\sqrt{%
u^{2}+a^{2}k^{2}}$ and then passed to polar coordinates in the plane $(u,ak)$%
. The representation (\ref{E2b}) is valid for all even values of $D$ and for
$D<7$ in the case of odd $D$. The shell-induced contribution (\ref{E2b})
depends on the variables $\eta $, $a$, $r$ in the form of the ratios $a/\eta
$ and $r/\eta $. This feature is a consequence of the maximal symmetry of dS
spacetime. Note that the combination $\alpha a/\eta $ is the proper radius
of the cylindrical shell and, hence, $a/\eta $ is the proper radius measured
in units of the dS curvature scale $\alpha $. Similarly, the ratio $r/\eta $
is the proper distance from the cylinder axis measured in units of $\alpha $.

As is seen from (\ref{E20}) and (\ref{E2b}), in the new representation of
the VEV the explicit knowledge of the eigenvalues $\gamma _{m,n}^{(\lambda
)} $ is not required. Another advantage is that we have explicitly separated
the part corresponding to the boundary-free dS spacetime. The presence of
the boundary does not change the local geometry for points outside the
shell. This means that at those points the divergences are the same in both
the problems, in the absence and in the presence of the cylindrical shell.
From here it follows that the divergences in (\ref{E2Dec}) are contained in
the part $\langle E^{2}\rangle _{\mathrm{dS}}$ only, whereas the
boundary-induced contribution $\left\langle E^{2}\right\rangle _{\mathrm{b}}$
is finite for points away from the boundary and the regularization,
implicitly assumed in the discussion above, can be safely removed in that
part. Hence, the renormalization is required for the boundary-free part
only. Note that the expression for the latter can be further simplified
after the summation over $m$ by using the standard result for the series
involving the square of the Bessel function.

Let us consider the properties of the boundary-induced contribution in the
VEV of the field squared. First of all, by taking into account that for $%
D\geqslant 3$ one has $G_{m}^{(0)}\left[ s,f(x)\right] >0$, $G_{m}^{(1)}%
\left[ s,f(x)\right] <0$, and the function $f_{D/2-2}(x)$ is positive for
the values of $D$ for which the representation (\ref{E2b}) is valid, from (%
\ref{E2b}) it follows that $\left\langle E^{2}\right\rangle _{\mathrm{b}}$
is always positive. The VEV of the electric field squared inside a
cylindrical shell in Minkowski spacetime is obtained by the limiting
transition $\alpha \rightarrow \infty $ for a fixed value of $t$. In this
limit one has $\eta \approx \alpha -t$ and, hence, $\eta $ is large. Passing
to a new integration variable $y=x/\eta $, we see that $\eta $ appears in
the argument of the function $f_{D/2-2}\left( ys\eta \right) $. By taking
into account that for large arguments one has $f_{D/2-2}\left( u\right)
\approx 1/u$, after the integration over $s$, we get $\lim_{\alpha
\rightarrow \infty }\left\langle E^{2}\right\rangle _{b}=\left\langle
E^{2}\right\rangle _{b}^{(\mathrm{M})}$, where%
\begin{equation}
\left\langle E^{2}\right\rangle _{\mathrm{b}}^{(\mathrm{M})}=\frac{4(4\pi
)^{(1-D)/2}}{\Gamma ((D+1)/2)}\sideset{}{'}{\sum}_{m=0}^{\infty
}\sum\limits_{\lambda =0,1}\int_{0}^{\infty }dx\,x^{D}\dfrac{K_{m}^{\left(
\lambda \right) }\left( ax\right) }{I_{m}^{\left( \lambda \right) }\left(
ax\right) }G_{(\mathrm{M})m}^{\left( \lambda \right) }\left[ I_{m}\left(
rx\right) \right] ,  \label{E2Mb}
\end{equation}%
with%
\begin{equation}
G_{(\mathrm{M})m}^{\left( \lambda \right) }\left[ f\left( x\right) \right]
=\left\{
\begin{array}{cc}
\left( D-2\right) \left[ f^{\prime 2}(x)+(m^{2}/x^{2}+2)f^{2}(x)\right] , &
\lambda =0, \\
-f^{\prime 2}(x)-m^{2}f^{2}(x)/x^{2}, & \lambda =1,%
\end{array}%
\right.  \label{GMmlam}
\end{equation}%
is the VEV of the electric field squared inside a cylindrical shell in the
Minkowski bulk.

In the special case of 4-dimensional dS spacetime one has $D=3$ and $%
f_{-1/2}(u)=1/u$. After the integration over $s$ in (\ref{E2b}), we find%
\begin{equation}
\left\langle E^{2}\right\rangle _{\mathrm{b}}=(\eta /\alpha
)^{4}\left\langle E^{2}\right\rangle _{\mathrm{b}}^{(\mathrm{M})},
\label{E2conf}
\end{equation}%
where $\left\langle E^{2}\right\rangle _{b}^{(\mathrm{M})}$ is given by (\ref%
{E2Mb}) with $D=3$. In this case, the VEV of the field squared is related to
the corresponding result in Minkowski spacetime by standard conformal
transformation with the conformal factor $(\eta /\alpha )^{4}$. This is a
direct consequence of the conformal invariance of the electromagnetic field
in $D=3$ spatial dimensions and of the conformal flatness of the background
geometry. The VEV\ of the electric field squared for a conducting
cylindrical shell coaxial with a cosmic string in 4-dimensional spacetime ($%
D=3$) is investigated in \cite{Beze07}. In the absence of planar angle
deficit, the corresponding expression is reduced to (\ref{E2Mb}) with $D=3$.
The corresponding Casimir-Polder forces were discussed in \cite{Beze11}.

On the axis of the shell, $r=0$, the only nonzero contribution to the
boundary-induced VEV comes from the terms in (\ref{E2b}) with $m=0,1$:
\begin{eqnarray}
\left\langle E^{2}\right\rangle _{\mathrm{b},r=0} &=&\frac{2^{4}A_{D}}{%
\alpha ^{D+1}}\left( \frac{\eta }{a}\right) ^{D+2}\int_{0}^{\infty
}dx\,x^{D+1}\int_{0}^{1}ds\,s\left( 1-s^{2}\right) ^{D/2-2}f_{D/2-2}\left(
xs\eta /a\right)   \notag \\
&&\times \left[ (\left( D-3\right) s^{2}+1)\dfrac{K_{0}\left( x\right) }{%
I_{0}\left( x\right) }+(1-s^{2})\dfrac{K_{1}\left( x\right) }{I_{1}\left(
x\right) }-s^{2}\dfrac{K_{1}^{\prime }\left( x\right) }{I_{1}^{\prime
}\left( x\right) }\right] .  \label{E2br0}
\end{eqnarray}%
The shell-induced VEV diverges on the cylindrical boundary. The surface
divergences in the VEVs of local physical observables are well-known in the
theory of the Casimir effect \cite{Most97} (see also \cite{Phil10} for
recent discussions). Near the shell the dominant contribution to the VEV\
comes from large values of $m$. For $m\neq 0$, introducing in (\ref{E2b}) a
new integration variable $y=x/m$, we use the uniform asymptotic expansions
for the modified Bessel functions with the order $m$ and the asymptotic for
the function $f_{D/2-2}\left( u\right) $ for large arguments (see, for
instance, \cite{Abra72}). To the leading order one finds%
\begin{equation}
\left\langle E^{2}\right\rangle _{\mathrm{b}}\approx \frac{3\left(
D-1\right) \Gamma ((D+1)/2)}{2^{D}\pi ^{(D-1)/2}[\alpha \left( a-r\right)
/\eta ]^{D+1}}.  \label{E2bnear}
\end{equation}%
Note that the combination $\alpha \left( a-r\right) /\eta $ is the proper
distance from the boundary. The leading term (\ref{E2bnear}) is obtained
from that for the cylindrical shell with the radius $a$ in Minkowski
spacetime by the replacement $(a-r)\rightarrow \alpha \left( a-r\right)
/\eta $. For points near the boundary the contribution of the modes with
small wavelengths dominate and at distances from the shell smaller than the
curvature radius of the dS spacetime the influence of the gravitational
field is small.

\begin{figure}[tbph]
\begin{center}
\epsfig{figure=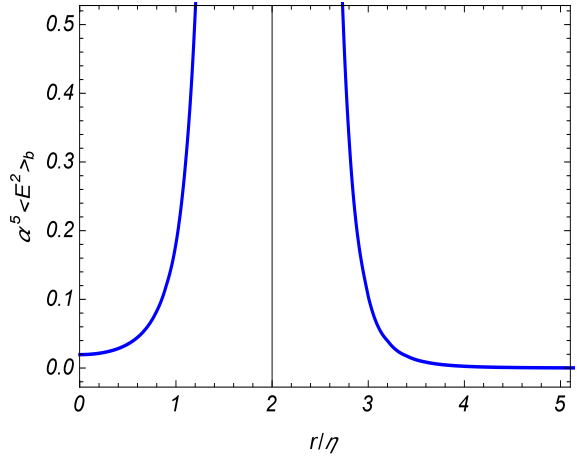,width=7.5cm,height=6.cm}
\end{center}
\caption{Shell-induced contribution in the VEV of the squared electric field
versus the proper distance from the shell axis for the $D=4$ model. For the
corresponding value of the shell radius we have taken $a/\protect\eta =2$.}
\label{fig1}
\end{figure}

For the numerical evaluations we have taken the model with $D=4$. In this
case for the function $f_{\nu }(y)$ in (\ref{E2b}) one has%
\begin{equation}
f_{\nu }(y)=2I_{\nu }(y)K_{\nu }(y),  \label{fD4}
\end{equation}%
and the integrals over $s$ are of the form%
\begin{equation}
\mathcal{I}_{n,\nu }(x)=\int_{0}^{1}ds\,s^{n}f_{\nu }(xs),  \label{Ical}
\end{equation}%
with $n=1,3$. These integrals are evaluated in appendix \ref{sec:App2}. For
the shell-induced part one gets
\begin{eqnarray}
\left\langle E^{2}\right\rangle _{\mathrm{b}} &=&\frac{2}{\pi ^{2}\alpha ^{5}%
}\sideset{}{'}{\sum}_{m=0}^{\infty }\sum\limits_{\lambda
=0,1}\int_{0}^{\infty }dx\,x^{5}\dfrac{K_{m}^{\left( \lambda \right) }\left(
xa/\eta \right) }{I_{m}^{\left( \lambda \right) }\left( xa/\eta \right) }
\notag \\
&&\times \left\{ \left[ I_{m}^{\prime 2}(y)+\left( m^{2}/y^{2}+1\right)
I_{m}^{2}(y)\right] \delta _{0\lambda }\mathcal{I}_{1,0}(x)\right.  \notag \\
&&-\left. \left[ I_{m}^{\prime 2}(y)+\left( m^{2}/y^{2}-\delta _{0\lambda
}\right) I_{m}^{2}(y)\right] \mathcal{I}_{3,0}(x)\right\} _{y=xr/\eta },
\label{E2bD4}
\end{eqnarray}%
with $\mathcal{I}_{1,0}(x)$ and $\mathcal{I}_{3,0}(x)$ given by (\ref{Int0}%
), (\ref{Int4}). In figure \ref{fig1}, this contribution is plotted versus
the proper distance from the shell axis, measured in units of $\alpha $. For
the proper radius of the shell, in the same units, we have taken $a/\eta =2$%
. The corresponding Casimir-Polder force is expressed in terms of the
derivative $\partial _{r}\left\langle E^{2}\right\rangle _{\mathrm{b}}$.
This force is directed toward the cylindrical shell.

\section{VEV of the energy-momentum tensor}

\label{sec:EMT}

Another important local characteristic of the vacuum state is the VEV of the
energy-momentum tensor. This VEV is evaluated by using the mode-sum formula%
\begin{equation}
\langle T_{\mu }^{\nu }\rangle =\frac{1}{16\pi }\delta _{\mu }^{\nu }g^{\rho
l}g^{\sigma m}\sum_{\beta }F_{(\beta )\rho \sigma }F_{(\beta )lm}^{\ast }-%
\frac{1}{4\pi }g^{\nu \kappa }g^{\rho \sigma }\sum_{\beta }F_{(\beta )\mu
\rho }F_{(\beta )\kappa \sigma }^{\ast },  \label{Tmunu}
\end{equation}%
where $F_{(\beta )\mu \rho }=\partial _{\mu }A_{(\beta )\rho }-\partial
_{\rho }A_{(\beta )\mu }$ is the field strength corresponding to the mode
functions (\ref{A1}) and (\ref{A2}). By using the expressions for the mode
functions, the mode-sums for the diagonal components of the VEV are
presented in the form (no summation over $i$)%
\begin{eqnarray}
\langle T_{i}^{i}\rangle  &=&\frac{2B_{D}\eta ^{D+2}}{\alpha ^{D+1}a^{2}}%
\sideset{}{'}{\sum}_{m=0}^{\infty }\int_{0}^{\infty
}dkk^{D-3}\sum\limits_{\lambda =0,1}\sum_{n=1}^{\infty }\gamma
_{m,n}^{(\lambda )}  \notag \\
&&\times T_{m}(\gamma _{m,n}^{(\lambda )})\sum_{l=0,1}t_{\lambda
l}^{(i)}[k,\gamma _{m,n}^{(\lambda )}/a,J_{m}(\gamma r)]L_{D/2-1-l}(\omega
_{m,n}^{(\lambda )}\eta ),  \label{Tii}
\end{eqnarray}%
where
\begin{equation}
B_{D}=\frac{2^{2-D}}{\pi ^{D/2+1}\Gamma (D/2-1)}.  \label{BD}
\end{equation}%
Here we have used the notations%
\begin{equation}
t_{\lambda l}^{(i)}[k,\gamma ,f(y)]=(a_{\lambda l}^{(i)}k^{2}+b_{\lambda
l}^{(i)}\gamma ^{2})Z_{m}^{(i)}[f(y)]+(\left( D-3\right) c_{\lambda
l}^{(i)}k^{2}+d_{\lambda l}^{(i)}\gamma ^{2})f^{2}(y),  \label{til}
\end{equation}%
and%
\begin{eqnarray}
Z_{m}^{(i)}[f(y)] &=&f^{\prime 2}(y)+m^{2}f^{2}(y)/y^{2},\;i=0,3,\ldots ,D,
\notag \\
Z_{m}^{(i)}[f(y)] &=&f^{\prime 2}(y)-m^{2}f^{2}(y)/y^{2},\;i=1,2.  \label{Zi}
\end{eqnarray}%
For the energy density the coefficients in (\ref{til}) are given by the
expressions%
\begin{eqnarray}
a_{\lambda l}^{(0)} &=&\left(
\begin{array}{cc}
1 & 1 \\
1 & 1%
\end{array}%
\right) ,\;b_{\lambda l}^{(0)}=\left(
\begin{array}{cc}
D-2 & 0 \\
0 & 1%
\end{array}%
\right) ,  \notag \\
c_{\lambda l}^{(0)} &=&\left(
\begin{array}{cc}
1 & 1 \\
0 & 0%
\end{array}%
\right) ,\;d_{\lambda l}^{(0)}=\left(
\begin{array}{cc}
0 & D-2 \\
1 & 0%
\end{array}%
\right) ,  \label{a0}
\end{eqnarray}%
where the rows and columns are numbered by $\lambda =0,1$ and $l=0,1$,
respectively. For the coefficients in the radial and azimuthal components
one has%
\begin{equation}
a_{\lambda l}^{(1)}=\left(
\begin{array}{cc}
-1 & 1 \\
1 & -1%
\end{array}%
\right) ,\;c_{\lambda l}^{(1)}=\left(
\begin{array}{cc}
1 & -1 \\
0 & 0%
\end{array}%
\right) ,  \label{a1}
\end{equation}%
and $a_{\lambda l}^{(2)}=-a_{\lambda l}^{(1)}$, $b_{\lambda
l}^{(2)}=-b_{\lambda l}^{(1)}=b_{\lambda l}^{(0)}$, $c_{\lambda
l}^{(2)}=c_{\lambda l}^{(1)}$, $d_{\lambda l}^{(1)}=d_{\lambda
l}^{(2)}=-d_{\lambda l}^{(0)}$. For the axial components, $i=3,\ldots ,D$,
we get%
\begin{eqnarray}
a_{\lambda l}^{(i)} &=&\frac{1}{D-2}\left(
\begin{array}{cc}
D-4 & 2-D \\
D-4 & 2-D%
\end{array}%
\right) ,\;b_{\lambda l}^{(i)}=\left(
\begin{array}{cc}
D-4 & 0 \\
0 & -1%
\end{array}%
\right) ,  \notag \\
c_{\lambda l}^{(i)} &=&\frac{1}{D-2}\left(
\begin{array}{cc}
D-6 & 4-D \\
0 & 0%
\end{array}%
\right) ,\;d_{\lambda l}^{(i)}=\left(
\begin{array}{cc}
0 & 4-D \\
1 & 0%
\end{array}%
\right) .  \label{ai}
\end{eqnarray}

In addition to the diagonal component, the VEV\ of the energy-momentum
tensor has a nonzero off-diagonal component%
\begin{eqnarray}
\langle T_{0}^{1}\rangle &=&\frac{2^{3-D}i\eta ^{D+2}}{\pi ^{D}\alpha
^{D+1}a^{3}}\sideset{}{'}{\sum}_{m=0}^{\infty }\int d\mathbf{k}%
\sum_{n=1}^{\infty }\sum_{\lambda =0,1}(-1)^{\lambda }N_{\lambda
}T_{m}(\gamma _{m,n}^{(\lambda )})\omega _{m,n}^{(\lambda )}\gamma
_{m,n}^{(\lambda )2}  \notag \\
&&\times K_{D/2-2}(\omega _{m,n}^{(\lambda )}\eta e^{-\frac{i\pi }{2}%
})K_{D/2-1}(\omega _{m,n}^{(\lambda )}\eta e^{\frac{i\pi }{2}})J_{m}(\gamma
_{m,n}^{(\lambda )}r/a)J_{m}^{\prime }(\gamma _{m,n}^{(\lambda )}r/a).
\label{T10}
\end{eqnarray}%
where $N_{0}=D-2$, $N_{1}=1$. This component describes energy flux along the
radial direction (along the direction normal to the boundary).

After the application of the summation formula (\ref{SumForm}), with the
function%
\begin{equation}
f(x)=xL_{D/2-1-l}(\eta \sqrt{x^{2}/a^{2}+k^{2}})t_{\lambda
l}^{(i)}[k,x/a,J_{m}(xr/a)],  \label{fxEMT}
\end{equation}%
to the series over $n$ in (\ref{Tii}), the VEV of the energy-momentum tensor
is presented in the form%
\begin{equation}
\langle T_{\mu }^{\nu }\rangle =\langle T_{\mu }^{\nu }\rangle _{\mathrm{dS}%
}+\langle T_{\mu }^{\nu }\rangle _{\mathrm{b}},  \label{Tdec}
\end{equation}%
where $\langle T_{\mu }^{\nu }\rangle _{\mathrm{dS}}$ is the corresponding
VEV in the boundary-free dS spacetime and the contribution $\langle T_{\mu
}^{\nu }\rangle _{b}$ is induced by the presence of the cylindrical shell.
The boundary-free contribution corresponds to the first integral in the
right-hand side of the formula (\ref{SumForm}) and the boundary-induced
contribution comes from the second integral. For points outside the
cylindrical shell the boundary-induced contribution in (\ref{Tdec}) is
finite and the renormalization is reduced to the one for the boundary-free
part. From the maximal symmetry of the Bunch-Davies vacuum state it follows
that the latter does not depend on the spacetime point and has the form $%
\langle T_{\mu }^{\nu }\rangle _{\mathrm{dS}}=\mathrm{const}\cdot \delta
_{\mu }^{\nu }$.

With the help of the transformations similar to those we have used for the
VEV of the field squared, for the boundary-induced parts in the diagonal
components we get (no summation over $i$)%
\begin{eqnarray}
\langle T_{i}^{i}\rangle _{\mathrm{b}} &=&\frac{B_{D}}{\alpha ^{D+1}}%
\sideset{}{'}{\sum}_{m=0}^{\infty }\sum\limits_{\lambda
=0,1}\int_{0}^{\infty }dx\,x^{D+1}\dfrac{K_{m}^{\left( \lambda \right)
}\left( ax/\eta \right) }{I_{m}^{\left( \lambda \right) }\left( ax/\eta
\right) }\int_{0}^{1}ds\,s  \notag \\
&&\times \left( 1-s^{2}\right) ^{D/2-2}\sum_{l=0,1}P_{\lambda
l}^{(i)}[s,I_{m}(xr/\eta )]f_{D/2-1-l}(xs),  \label{Tiib}
\end{eqnarray}%
where we have defined the functions%
\begin{equation}
P_{\lambda l}^{(i)}[s,f(y)]=(A_{\lambda l}^{(i)}+B_{\lambda
l}^{(i)}s^{2})Z_{m}^{(i)}[f(y)]+(C_{\lambda l}^{(i)}+\left( D-3\right)
D_{\lambda l}^{(i)}s^{2})f^{2}(y),  \label{Pi}
\end{equation}%
with $Z_{m}^{(i)}[f(y)]$ given by (\ref{Zi}). For the energy density the
coefficients in (\ref{Pi}) are given by the expressions%
\begin{equation}
A_{\lambda l}^{(0)}=C_{\lambda l}^{(0)}=\left(
\begin{array}{cc}
3-D & 1 \\
1 & 0%
\end{array}%
\right) ,\;B_{\lambda l}^{(0)}=-\left(
\begin{array}{cc}
1 & 1 \\
1 & 1%
\end{array}%
\right) ,\;D_{\lambda l}^{(0)}=\left(
\begin{array}{cc}
1 & 1 \\
0 & 0%
\end{array}%
\right) .\;  \label{A0}
\end{equation}%
For the radial and azimuthal stresses one has%
\begin{eqnarray}
A_{\lambda l}^{(i)} &=&\left( -1\right) ^{i}C_{\lambda l}^{(i)}=-\left(
-1\right) ^{i}\left(
\begin{array}{cc}
D-3 & 1 \\
1 & 0%
\end{array}%
\right) ,  \notag \\
B_{\lambda l}^{(i)} &=&\left( -1\right) ^{i}\left(
\begin{array}{cc}
-1 & 1 \\
1 & -1%
\end{array}%
\right) ,\;D_{\lambda l}^{(i)}=\left(
\begin{array}{cc}
1 & -1 \\
0 & 0%
\end{array}%
\right) ,\;  \label{Ai1}
\end{eqnarray}%
with $i=1,2$. And finally, for the axial stresses ($i=3,\ldots ,D$) we get:%
\begin{eqnarray}
A_{\lambda l}^{(i)} &=&\frac{1}{D-2}\left(
\begin{array}{cc}
(D-4)\left( 3-D\right) & 2-D \\
D-4 & 0%
\end{array}%
\right) ,\;B_{\lambda l}^{(i)}=\frac{1}{D-2}\left(
\begin{array}{cc}
4-D & D-2 \\
4-D & D-2%
\end{array}%
\right) ,  \notag \\
C_{\lambda l}^{(i)} &=&\frac{1}{D-2}\left(
\begin{array}{cc}
\left( 3-D\right) \left( D-6\right) & 4-D \\
D-2 & 0%
\end{array}%
\right) ,\;D_{\lambda l}^{(i)}=\frac{1}{D-2}\left(
\begin{array}{cc}
D-6 & 4-D \\
0 & 0%
\end{array}%
\right) .\;  \label{Ai2}
\end{eqnarray}%
Note that, unlike to the case of the Minkowski bulk (see below), for the dS
bulk the axial stresses do not coincide with the energy density.

For the off-diagonal component (\ref{T10}), the contribution of the first
integral in (\ref{SumForm}) to the VEV vanishes. This directly follows from
the relations $\sum_{m=0}^{\prime \infty }J_{m}(x)J_{m}^{\prime
}(x)=(1/2)\partial _{x}\sum_{m=0}^{\prime \infty }J_{m}^{2}(x)$ and $%
\sum_{m=0}^{\prime \infty }J_{m}^{2}(x)=1/2$. The nonzero part is induced by
the presence of the shell and is given by the expression%
\begin{eqnarray}
\langle T_{0}^{1}\rangle _{\mathrm{b}} &=&-\frac{2B_{D}}{\alpha ^{D+1}}%
\sideset{}{'}{\sum}_{m=0}^{\infty }\int_{0}^{\infty }dx\,x^{D+1}\left[ (D-2)%
\frac{K_{m}(xa/\eta )}{I_{m}(xa/\eta )}-\frac{K_{m}^{\prime }(xa/\eta )}{%
I_{m}^{\prime }(xa/\eta )}\right] I_{m}(xr/\eta )I_{m}^{\prime }(xr/\eta )
\notag \\
&&\times \int_{0}^{1}ds\,s^{2}\left( 1-s^{2}\right) ^{D/2-2}\left[
K_{D/2-1}\left( y\right) I_{2-D/2}\left( y\right) -K_{D/2-2}\left( y\right)
I_{D/2-1}\left( y\right) \right] _{y=xs}.  \label{T01b}
\end{eqnarray}%
In the special case $D=3$ the off-diagonal component vanishes. For other
values of $D$, for which the representation (\ref{T01b}) is valid both the
functions in the square brackets are positive and, hence, $\langle
T_{0}^{1}\rangle _{\mathrm{b}}<0$ for $0<r<a$. On the axis, the energy flux
vanishes, $\langle T_{0}^{1}\rangle _{\mathrm{b},r=0}=0$. Similar to the
case of the field squared, the boundary-induced VEVs (\ref{Tiib}) and (\ref%
{T01b}) depend on $\eta $, $a$, $r$ in the form of the ratios $a/\eta $ and $%
r/\eta $.

With the expressions (\ref{Tiib}) and (\ref{T01b}), we can see that the
boundary-induced contributions obey the covariant continuity equation $%
\nabla _{\nu }\langle T_{\mu }^{\nu }\rangle _{b}=0$. For the geometry at
hand, this equation is reduced to the following relations between the VEVs:%
\begin{eqnarray}
\left( \partial _{\eta }-\frac{D+1}{\eta }\right) \langle T_{0}^{0}\rangle _{%
\mathrm{b}} &=&\left( \partial _{r}+\frac{1}{r}\right) \langle
T_{0}^{1}\rangle _{\mathrm{b}}-\frac{1}{\eta }\langle T_{k}^{k}\rangle _{%
\mathrm{b}},  \notag \\
\left( \partial _{\eta }-\frac{D+1}{\eta }\right) \langle T_{0}^{1}\rangle _{%
\mathrm{b}} &=&\frac{1}{r}\langle T_{2}^{2}\rangle _{\mathrm{b}}-\left(
\partial _{r}+\frac{1}{r}\right) \langle T_{1}^{1}\rangle _{\mathrm{b}}.
\label{Conteq}
\end{eqnarray}%
Let us denote by $\mathcal{E}_{\mathrm{b},r\leqslant r_{0}}$ the
shell-induced contribution in the vacuum energy in the region $r\leqslant
r_{0}<a$, per unit coordinate lengths along the directions $z^{3},\ldots
,z^{D}$:
\begin{equation}
\mathcal{E}_{\mathrm{b},r\leqslant r_{0}}=2\pi \left( \alpha /\eta \right)
^{D}\int_{0}^{r_{0}}dr\,r\langle T_{0}^{0}\rangle _{\mathrm{b}}.
\label{Ebin}
\end{equation}%
By taking into account the first equation in (\ref{Conteq}), the
corresponding derivative with respect to the synchronous time coordinate $t$
is expressed as (see also \cite{Saha15})%
\begin{equation}
\partial _{t}\mathcal{E}_{\mathrm{b},r\leqslant r_{0}}=\frac{2\pi }{\alpha }%
\left( \alpha /\eta \right) ^{D}\int_{0}^{r_{0}}dr\,r\sum_{l=1}^{D}\langle
T_{l}^{l}\rangle _{\mathrm{b}}-2\pi r_{0}\left( \alpha /\eta \right)
^{D-1}\langle T_{0}^{1}\rangle _{\mathrm{b},r=r_{0}}.  \label{Eder}
\end{equation}%
This relation shows that $\langle T_{0}^{1}\rangle _{b}$ is the energy flux
per unit proper surface area. The quantity $-\langle T_{l}^{l}\rangle _{b}$
is the shell-induced contribution to the vacuum pressure along the $l$-th
direction and the first term in the right-hand side of (\ref{Eder}) is the
work done by the surrounding on the selected volume. The last term in (\ref%
{Eder}) is the energy flux through the surface $r=r_{0}$. Inside the shell
one has $\langle T_{0}^{1}\rangle _{\mathrm{b}}<0$ and the flux is directed
from the shell to the axis $r=0$.

Let us discuss special cases of the general expressions for the VEVs of the
energy-momentum tensor components. First we consider the Minkowskian limit,
corresponding to $\alpha \rightarrow \infty $. Introducing in (\ref{Tiib}) a
new integration variable $y=x/\eta $, we see that the argument of the
functions $f_{\nu }(u)$ is large and we can use the asymptotic expression $%
f_{\nu }(u)\approx 1/u$. After the integration over $s$, to the leading
order we get $\langle T_{i}^{i}\rangle _{b}\approx \langle T_{i}^{i}\rangle
_{b}^{(\mathrm{M})}$, where for the VEVs on the Minkowski bulk one has (no
summation over $i$)%
\begin{equation}
\langle T_{i}^{i}\rangle _{\mathrm{b}}^{(\mathrm{M})}=\frac{2(4\pi
)^{-(D+1)/2}}{\Gamma ((D+1)/2)}\sideset{}{'}{\sum}_{m=0}^{\infty
}\sum\limits_{\lambda =0,1}\int_{0}^{\infty }dxx^{D}\dfrac{K_{m}^{\left(
\lambda \right) }\left( ax\right) }{I_{m}^{\left( \lambda \right) }\left(
ax\right) }\left\{ A_{\lambda }^{(i)}Z_{m}^{(i)}[I_{m}(rx)]+B_{\lambda
}^{(i)}I_{m}^{2}(rx)\right\} ,  \label{TiiM}
\end{equation}%
with the coefficients%
\begin{eqnarray}
A_{0}^{(0)} &=&\left( 2-D\right) (D-3),\;A_{1}^{(0)}=D-3,  \notag \\
B_{0}^{(0)} &=&(2-D)\left( D-5\right) ,\;B_{1}^{(0)}=D-1,  \notag \\
A_{\lambda }^{(l)} &=&\left( -1\right) ^{l}B_{\lambda
}^{(l)},\;B_{0}^{(l)}=(2-D)\left( D-1\right) ,\;B_{1}^{(l)}=1-D,
\label{A00M}
\end{eqnarray}%
for $l=1,2$. For the stresses along the directions $i=3,\ldots ,D$ we have $%
\langle T_{i}^{i}\rangle _{b}^{(\mathrm{M})}=\langle T_{0}^{0}\rangle _{b}^{(%
\mathrm{M})}$. In the special case $D=3$, from (\ref{TiiM}) we obtain the
results previously derived in \cite{Saha88}. As is seen, for the Minkowski
bulk the axial stresses are equal to the energy density. This result could
be directly obtained on the base of the invariance of the problem with
respect to the Lorentz boosts along the directions of the axis $x^{i}$, $%
i=3,\ldots ,D$. The off-diagonal component vanishes in the Minkowskian
limit. For the leading term in the corresponding asymptotic expansion from (%
\ref{T01b}) we find%
\begin{equation}
\langle T_{0}^{1}\rangle _{\mathrm{b}}\approx \frac{3-D}{\alpha }\frac{%
2^{1-D}\pi ^{-(D+1)/2}}{\Gamma ((D-1)/2)}\sideset{}{'}{\sum}_{m=0}^{\infty
}\int_{0}^{\infty }dx\,x^{D-1}\left[ (D-2)\frac{K_{m}(ax)}{I_{m}(ax)}-\frac{%
K_{m}^{\prime }(ax)}{I_{m}^{\prime }(ax)}\right] I_{m}(xr)I_{m}^{\prime
}(xr).  \label{T01bM}
\end{equation}

In the special case $D=3$ the off-diagonal component of the vacuum
energy-momentum tensor vanishes, $\langle T_{0}^{1}\rangle _{\mathrm{b}}=0$,
and the diagonal components are connected to the corresponding quantities
for a cylindrical shell in the Minkowski bulk by the relation (no summation
over $i$)%
\begin{equation}
\langle T_{i}^{i}\rangle _{\mathrm{b}}=(\eta /\alpha )^{4}\langle
T_{i}^{i}\rangle _{\mathrm{b}}^{(\mathrm{M})}.  \label{TiiD3}
\end{equation}%
In this special case $A_{\lambda }^{(i)}=0$ for $i=0,3$, and $%
B_{0}^{(i)}=B_{1}^{(i)}$ for all $i$. Note that the Casimir self-stress for
an infinite perfectly conducting cylindrical shell in background of
4-dimensional Minkowski spacetime has been evaluated in \cite{Dera81} on the
base of a Green's function technique. The corresponding Casimir energy was
investigated by using the zeta function technique in \cite{Gosd98} and the
mode-by-mode summation method in \cite{Milt99}. The geometry of a
cylindrical shell coaxial with a cosmic string was considered in \cite%
{Beze07,Brev95}.

Near the cylindrical shell, the asymptotic expressions for the components of
the energy-momentum tensor are found in the way similar to that for the VEV
of the electric field squared, by using the uniform asymptotic expansions
for the modified Bessel functions. The leading terms are given by the
expressions (no summation over $i$)%
\begin{equation}
\langle T_{i}^{i}\rangle _{\mathrm{b}}\approx -\frac{\left( D-1\right)
(D-3)\Gamma ((D+1)/2)}{2(4\pi )^{(D+1)/2}[\alpha \left( a-r\right) /\eta
]^{D+1}},  \label{Tiinear}
\end{equation}%
for $i=0,2,\ldots ,D$, and%
\begin{equation}
\langle T_{0}^{1}\rangle _{\mathrm{b}}\approx \frac{a-r}{\eta }\langle
T_{0}^{0}\rangle _{b},\;\langle T_{1}^{1}\rangle _{\mathrm{b}}\approx \frac{%
a-r}{Da}\langle T_{0}^{0}\rangle _{b}.  \label{T10near}
\end{equation}%
The leading terms for the diagonal components coincide with those for a
cylindrical shell in Minkowski bulk with the distance from the shell
replaced by the proper distance $\alpha \left( a-r\right) /\eta $. In the
special case $D=3$ the leading terms vanish. The latter feature is related
to the conformal invariance of the electromagnetic field in $D=3$.

In the special case $D=4$, the integrals over $s$ in (\ref{Tiib}) are
evaluated in appendix \ref{sec:App2}. The expression for the diagonal
components takes the form (no summation over $i$)%
\begin{eqnarray}
\langle T_{i}^{i}\rangle _{\mathrm{b}} &=&\frac{\alpha ^{-5}}{4\pi ^{3}}%
\sideset{}{'}{\sum}_{m=0}^{\infty }\sum\limits_{\lambda
=0,1}\int_{0}^{\infty }dxx^{5}\dfrac{K_{m}^{\left( \lambda \right) }\left(
ax/\eta \right) }{I_{m}^{\left( \lambda \right) }\left( ax/\eta \right) }
\notag \\
&&\times \sum_{l=0,1}\left\{ (A_{\lambda l}^{(i)}\mathcal{I}%
_{1,1-l}(x)+B_{\lambda l}^{(i)}\mathcal{I}_{3,1-l}(x))Z_{m}^{(i)}[I_{m}(xr/%
\eta )]\right.   \notag \\
&&\left. +(C_{\lambda l}^{(i)}\mathcal{I}_{1,1-l}(x)+D_{\lambda l}^{(i)}%
\mathcal{I}_{3,1-l}(x))I_{m}^{2}(xr/\eta )\right\} ,  \label{TiiD4}
\end{eqnarray}%
where the coefficients are given by (\ref{A0}), (\ref{Ai1}) and (\ref{Ai2})
with $D=4$. In the expression (\ref{T01b}) for the energy flux, with $D=4$,
the integral over $s$ is evaluated by using the formulas (\ref{Int11}) and (%
\ref{Int12}). This leads to the following result:
\begin{eqnarray}
\langle T_{0}^{1}\rangle _{\mathrm{b}} &=&-\frac{\alpha ^{-5}}{2\pi ^{3}}%
\sideset{}{'}{\sum}_{m=0}^{\infty }\int_{0}^{\infty }dx\,x^{4}\left[ 2\frac{%
K_{m}(ax/\eta )}{I_{m}(ax/\eta )}-\frac{K_{m}^{\prime }(ax/\eta )}{%
I_{m}^{\prime }(ax/\eta )}\right]   \notag \\
&&\times I_{m}(xr/\eta )I_{m}^{\prime }(xr/\eta )I_{1}\left( x\right)
K_{1}\left( x\right) .  \label{T01D4}
\end{eqnarray}

\begin{figure}[tbph]
\begin{center}
\begin{tabular}{cc}
\epsfig{figure=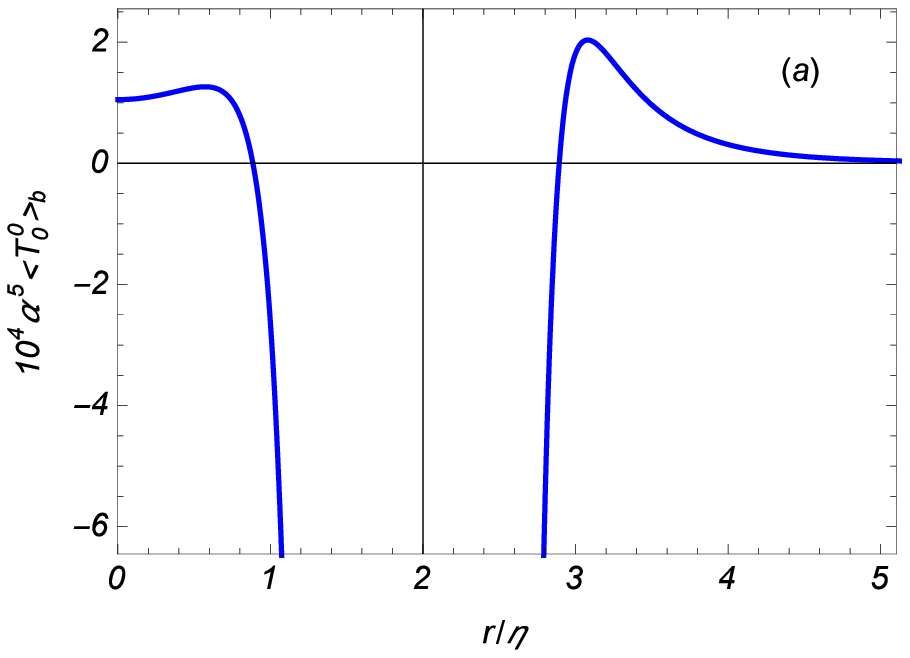,width=7.cm,height=6.cm} & \quad %
\epsfig{figure=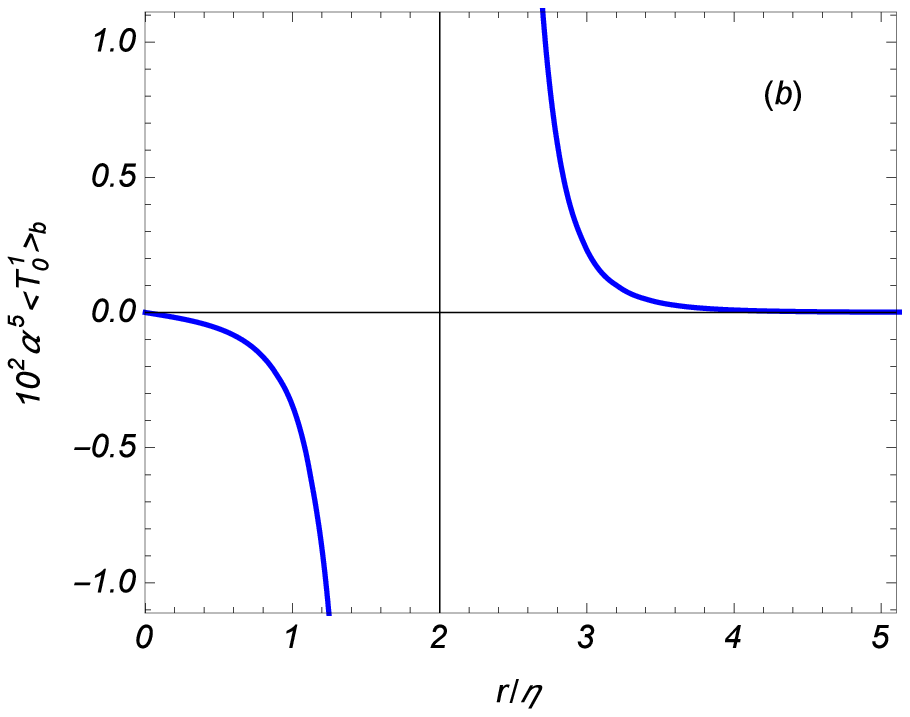,width=7.cm,height=6.cm}%
\end{tabular}%
\end{center}
\caption{Shell-induced contribution in the VEV of the energy density (panel
(a)) and the energy flux (panel (b)) as functions of the ratio $r/\protect%
\eta $ in the $D=4$ model. The graphs are plotted for $a/\protect\eta =2$.}
\label{fig2}
\end{figure}

In figure \ref{fig2} we have plotted the boundary-induced contribution in
the energy density and the energy flux as functions of the proper distance
from the shell axis, measured in units of the dS curvature scale $\alpha $.
For the corresponding value of the shell radius we have taken $a/\eta =2$.
As is seen, in the interior region the energy density is negative near the
shell and positive near the axis of the shell. The energy flux is negative
inside the shell. This means that the energy flux is directed from the
shell. The corresponding energy density in the Minkowski bulk is negative
everywhere.

\section{Exterior region}

\label{sec:Ext}

In the region outside the shell, $r>a$, the radial functions $C_{m}(x)$ in (%
\ref{A1}) and (\ref{A2}) are linear combinations of the Bessel and Neumann
functions. The relative coefficients in the linear combinations are
determined from the boundary condition (\ref{BC1}) on the cylindrical
surface $r=a$. In this way we can see that
\begin{equation}
C_{m}(\gamma r)=g_{m}^{(\lambda )}(\gamma a,\gamma r)=J_{m}(\gamma
r)Y_{m}^{(\lambda )}(\gamma a)-Y_{m}(\gamma r)J_{m}^{(\lambda )}(\gamma a),
\label{RadOut}
\end{equation}%
for the modes $\lambda =0,1$. Now, in the normalization condition (\ref%
{NormCond}) the integration over the radial coordinate goes over the region $%
a\leqslant r<\infty $ and in the right-hand side the delta symbol for the
quantum number $\gamma $ is understood as the Dirac delta function $\delta
(\gamma -\gamma ^{\prime })$. As the normalization integral diverges for $%
\gamma ^{\prime }=\gamma $, the main contribution to the integral comes from
large values of $r$. By making use of the asymptotic formulas for the Bessel
and Neumann functions with large arguments, for the normalization
coefficients in (\ref{A1}) and (\ref{A2}) one gets
\begin{equation}
\left\vert c_{\beta }\right\vert ^{2}=\frac{[J_{m}^{(\lambda )2}(\gamma
a)+Y_{m}^{(\lambda )2}(\gamma a)]^{-1}}{4\left( 2\pi \alpha \right)
^{D-3}\gamma }.  \label{Cout}
\end{equation}%
Similar to the case of the interior region, we consider the VEVs of the
field squared and energy-momentum tensor separately.

\subsection{VEV of the field squared}

For the VEV of the electric field squared, by using the mode-sum formula (%
\ref{E2mode}), one finds
\begin{eqnarray}
\langle E^{2}\rangle &=&\frac{2^{5}A_{D}\eta ^{D+2}}{\alpha ^{D+1}}\ %
\sideset{}{'}{\sum}_{m=0}^{\infty }\int_{0}^{\infty
}dkk^{D-3}\int_{0}^{\infty }d\gamma \,  \notag \\
&&\times \gamma ^{3}\sum\limits_{\lambda =0,1}\frac{G_{m}^{(\lambda
)}[k,g_{m}^{(\lambda )}(\gamma a,\gamma r)]}{J_{m}^{(\lambda )2}(\gamma
a)+Y_{m}^{(\lambda )2}(\gamma a)}L_{D/2-2}(\omega \eta ),  \label{E2out}
\end{eqnarray}%
with the same notations as in (\ref{E2}). With the help of the identity%
\begin{equation}
\frac{G_{m}^{(\lambda )}[k,g_{m}^{(\lambda )}(\gamma a,\gamma r)]}{%
J_{m}^{(\lambda )2}(\gamma a)+Y_{m}^{(\lambda )2}(\gamma a)}=G_{m}^{(\lambda
)}[J_{m}(\gamma r)]-\frac{1}{2}\sum\limits_{j=0,1}\frac{J_{m}^{\left(
\lambda \right) }\left( \gamma a\right) }{H_{m}^{\left( j\right) \left(
\lambda \right) }\left( \gamma a\right) }G_{m}^{(\lambda )}[k,H_{m}^{\left(
j\right) }(\gamma r)],  \label{Ident}
\end{equation}%
the VEV is presented in the decomposed form (\ref{E2Dec}) where the
shell-induced contribution is given by the expression%
\begin{eqnarray}
\langle E^{2}\rangle _{\mathrm{b}} &=&-\frac{2^{4}A_{D}\eta ^{D+2}}{\alpha
^{D+1}}\sideset{}{'}{\sum}_{m=0}^{\infty }\int_{0}^{\infty }dkk^{D-3}\
\sum\limits_{j=0,1}  \notag \\
&&\times \sum\limits_{\lambda =0,1}\int_{0}^{\infty }d\gamma \gamma ^{3}%
\frac{J_{m}^{\left( \lambda \right) }\left( \gamma a\right) }{H_{m}^{\left(
j\right) \left( \lambda \right) }\left( \gamma a\right) }L_{D/2-2}(\omega
\eta ).  \label{E2bout}
\end{eqnarray}

For the further transformation, in (\ref{E2bout}), we rotate the contour of
integration in the complex plane $\gamma $ by the angle $\pi /2$ for the
term with $j=1$ and by the angle $-\pi /2$ for $j=2$. Introducing the
modified Bessel functions, the expression (\ref{E2bout}) takes the form
\begin{eqnarray}
\langle E^{2}\rangle _{\mathrm{b}} &=&\frac{2^{5}A_{D}}{\alpha ^{D+1}}%
\sideset{}{'}{\sum}_{m=0}^{\infty }\sum\limits_{\lambda
=0,1}\int_{0}^{\infty }dx\,x^{D+1}\dfrac{I_{m}^{\left( \lambda \right)
}\left( ax/\eta \right) }{K_{m}^{\left( \lambda \right) }\left( ax/\eta
\right) }  \notag \\
&&\times \int_{0}^{1}dss\left( 1-s^{2}\right) ^{D/2-2}G_{m}^{\left( \lambda
\right) }\left[ s,K_{m}\left( xr/\eta \right) \right] f_{D/2-2}\left(
xs\right) ,  \label{E2bout2}
\end{eqnarray}%
where the functions $G_{m}^{(\lambda )}\left[ s,f(x)\right] $ are defined by
the formulae (\ref{Gb}). Comparing this result with (\ref{E2b}), we see that
the expressions for the interior and exterior regions are related by the
interchange $I_{m}\rightleftarrows K_{m}$. In particular, the result in the
Minkowskian limit is obtained from (\ref{E2Mb}) by the same replacements. In
the special case $D=3$ the electromagnetic field is conformally invariant
and we have the relation (\ref{E2conf}). Similar to the case of the interior
region, the shell-induced contribution (\ref{E2bout}) in the VEV of the
electric field squared is positive. For points near the shell, the leading
term in the asymptotic expansion over the distance from the boundary is
obtained from (\ref{E2bnear}) by the replacement $a-r\rightarrow r-a$.

At large proper distances from the shell compared with the dS curvature
radius, we have $r/\eta \gg 1$. Introducing in (\ref{E2bout}) a new
integration variable $y=xr/\eta $ and assuming that $r\gg a$, we use the
expansions of the functions $I_{m}^{\left( \lambda \right) }\left(
ya/r\right) /K_{m}^{\left( \lambda \right) }\left( ya/r\right) $ and $%
f_{D/2-2}\left( ys\eta /r\right) $ for small values of the arguments. The
leading contribution comes from the term with $\lambda =0$ and $m=0$. For
even $D>4$ one gets
\begin{equation}
\langle E^{2}\rangle _{\mathrm{b}}\approx \frac{4\left( 4D^{2}-3D-4\right)
\Gamma ^{3}(D/2+1)}{\pi ^{D/2}D(D-4)\Gamma (D+2)(\alpha r/\eta )^{D+2}}\frac{%
\alpha }{\ln (r/a)}.  \label{E2bfar}
\end{equation}%
In the case $D=4$ the leading term is given by
\begin{equation}
\langle E^{2}\rangle _{\mathrm{b}}\approx \frac{16\pi ^{-2}\alpha (\alpha
r/\eta )^{-6}}{5\ln (r/a)\ln (r/\eta )}.  \label{E2bfarD4}
\end{equation}%
For $D=3$ and $D=5$ we find%
\begin{eqnarray}
\langle E^{2}\rangle _{\mathrm{b}} &\approx &\frac{2(\alpha r/\eta )^{-4}}{%
3\pi \ln (r/a)},\;D=3,  \notag \\
\langle E^{2}\rangle _{\mathrm{b}} &\approx &\frac{7(\alpha r/\eta )^{-6}}{%
10\pi ^{2}\ln (r/a)},\;D=5.  \label{E2bfarD35}
\end{eqnarray}%
At large distances, the total VEV is dominated by the boundary-free part $%
\langle E^{2}\rangle _{\mathrm{dS}}$. For a cylindrical shell in the
Minkowski bulk, at large distances, $r\gg a$, one has the following
asymptotic behavior:%
\begin{equation}
\left\langle E^{2}\right\rangle _{\mathrm{b}}^{(\mathrm{M})}\approx \frac{%
\left( D-2\right) \left( 3D-1\right) }{\pi ^{(D-1)/2}r^{D+1}\ln (r/a)}\frac{%
\Gamma ^{3}((D+1)/2)}{\left( D-1\right) \Gamma (D+1)},  \label{E2bMfar}
\end{equation}%
for all values $D\geqslant 3$. Note that for $D=3,5$ the leading term (\ref%
{E2bMfar}) is obtained from (\ref{E2bfarD35}) by the replacement of the
proper distance from the axis, $\alpha r/\eta \rightarrow r$.

For the special case $D=4$, the shell-induced contribution in the VEV of the
squared electric field, for the exterior region, is presented in figure \ref%
{fig1} as a function of the ratio $r/\eta $. Similar to the region inside
the shell, the corresponding Casimir-Polder force is directed toward the
shell.

\subsection{Energy-momentum tensor}

Now let us consider the VEV of the energy-momentum tensor outside the
cylindrical shell. By using the mode-sum formula (\ref{Tmunu}) with the
exterior modes, for the VEV of the diagonal components we get the following
representation (no summation over $i$)%
\begin{eqnarray}
\langle T_{i}^{i}\rangle &=&\frac{B_{D}\eta ^{D+2}}{\alpha ^{D+1}}%
\sideset{}{'}{\sum}_{m=0}^{\infty }\int_{0}^{\infty
}dkk^{D-3}\sum\limits_{\lambda =0,1}\int_{0}^{\infty }d\gamma \gamma  \notag
\\
&&\times \sum_{l=1,2}\frac{t_{\lambda l}^{(i)}[k,\gamma ,g_{m}^{(\lambda
)}(\gamma a,\gamma r)]}{J_{m}^{(\lambda )2}(\gamma a)+Y_{m}^{(\lambda
)2}(\gamma a)}L_{D/2-1-l}(\omega \eta ),  \label{Tiiext}
\end{eqnarray}%
with the notation (\ref{til}). The expression for the off-diagonal component
has the form%
\begin{eqnarray}
\langle T_{0}^{1}\rangle &=&\frac{2iB_{D}\eta ^{D+2}}{\alpha ^{D+1}}%
\sideset{}{'}{\sum}_{m=0}^{\infty }\int_{0}^{\infty
}dkk^{D-3}\int_{0}^{\infty }d\gamma \gamma \sum_{\lambda =0,1}(-1)^{\lambda
}N_{\lambda }\omega \gamma ^{2}  \notag \\
&&\times K_{D/2-2}(\omega \eta e^{-\frac{i\pi }{2}})K_{D/2-1}(\omega \eta e^{%
\frac{i\pi }{2}})\frac{g_{m}^{(\lambda )}(\gamma a,\gamma r)g_{m}^{(\lambda
)\prime }(\gamma a,\gamma r)}{J_{m}^{(\lambda )2}(\gamma a)+Y_{m}^{(\lambda
)2}(\gamma a)},  \label{T10ext}
\end{eqnarray}%
where $g_{m}^{(\lambda )\prime }(x,y)=\partial _{y}g_{m}^{(\lambda )}(x,y)$.

Further transformation of the VEVs is similar to that we have used for the
field squared. In the case of the diagonal components we employ the relation%
\begin{equation}
\frac{t_{\lambda l}^{(i)}[k,\gamma ,g_{m}^{(\lambda )}(\gamma a,\gamma r)]}{%
J_{m}^{(\lambda )2}(\gamma a)+Y_{m}^{(\lambda )2}(\gamma a)}=t_{\lambda
l}^{(i)}[k,\gamma ,J_{m}(\gamma r)]-\frac{1}{2}\sum\limits_{j=1,2}\frac{%
J_{m}^{\left( \lambda \right) }\left( \gamma a\right) }{H_{m}^{\left(
j\right) \left( \lambda \right) }\left( \gamma a\right) }t_{\lambda
l}^{(i)}[k,\gamma ,H_{m}^{\left( j\right) }(\gamma r)],  \label{Ident2}
\end{equation}%
The part with the first term in the right-hand side coincides with the VEV
in the boundary-free dS spacetime. In the part corresponding to the last
term in (\ref{Ident2}) we rotate the integration contours by the angle $\pi
/2$ for the term with $j=1$ and by the angle $-\pi /2$ for $j=2$. In this
way for the boundary-induced contributions in the VEVs of the diagonal
components we find (no summation over $i$)%
\begin{eqnarray}
\langle T_{i}^{i}\rangle _{\mathrm{b}} &=&\frac{B_{D}}{\alpha ^{D+1}}%
\sideset{}{'}{\sum}_{m=0}^{\infty }\sum\limits_{\lambda
=0,1}\int_{0}^{\infty }dxx^{D+1}\dfrac{I_{m}^{\left( \lambda \right) }\left(
ax/\eta \right) }{K_{m}^{\left( \lambda \right) }\left( ax/\eta \right) }%
\int_{0}^{1}ds\,s  \notag \\
&&\times \left( 1-s^{2}\right) ^{D/2-2}\sum_{l=0,1}P_{\lambda
l}^{(i)}[s,K_{m}(xr/\eta )]f_{D/2-1-l}(xs).  \label{Tiibout}
\end{eqnarray}%
In a similar way, the expression for the VEV of the off-diagonal component
is presented as
\begin{eqnarray}
\langle T_{0}^{1}\rangle _{\mathrm{b}} &=&-\frac{2B_{D}}{\alpha ^{D+1}}%
\sideset{}{'}{\sum}_{m=0}^{\infty }\int_{0}^{\infty }dx\,x^{D+1}\left[ (D-2)%
\frac{I_{m}(xa/\eta )}{K_{m}(xa/\eta )}-\frac{I_{m}^{\prime }(xa/\eta )}{%
K_{m}^{\prime }(xa/\eta )}\right] K_{m}(xr/\eta )K_{m}^{\prime }(xr/\eta )
\notag \\
&&\times \int_{0}^{1}ds\,s^{2}\left( 1-s^{2}\right) ^{D/2-2}\left[
K_{D/2-1}\left( y\right) I_{2-D/2}\left( y\right) -K_{D/2-2}\left( y\right)
I_{D/2-1}\left( y\right) \right] _{y=xs}.  \label{T10bout}
\end{eqnarray}%
For $D\geqslant 4$, in the range of validity of this representation one has $%
\langle T_{0}^{1}\rangle _{\mathrm{b}}>0$. The corresponding energy flux is
directed from the cylindrical shell to the infinity. Again, we can see that
the shell-induced contributions obey the relations (\ref{Conteq}).

The VEVs in the Minkowskian limit, $\alpha \rightarrow \infty $, are
obtained in the way similar to that for the interior region. In this limit
the energy flux vanishes and the corresponding expressions for the diagonal
components are obtained from (\ref{TiiM}) by the replacements $%
I_{m}\rightleftarrows K_{m}$. In the special case $D=3$, the VEVs in the dS
and Minkowski bulks are connected by the conformal relation (\ref{TiiD3}).

For points near the shell, the leading terms in the asymptotic expansions
over the distance from the boundary for the components $\langle
T_{i}^{i}\rangle _{\mathrm{b}}$ with $i\neq 1$ are given by (\ref{Tiinear})
with the replacement $a-r\rightarrow r-a$. For the normal stress and the
energy flux we have the relations (\ref{T10near}). Hence, near the shell and
for $D>3$, the boundary-induced contribution in the energy density has the
same sign in the exterior and interior regions, whereas the normal stress
and the energy flux have opposite signs.

Let us consider the asymptotics of the boundary-induced contributions at
large distances from the shell. Introducing in (\ref{Tiibout}) and (\ref%
{T10bout}) a new integration variable $y=xr$, we see that for $r\gg a,\eta $
the arguments of the functions $I_{m}^{\left( \lambda \right) }$, $%
K_{m}^{\left( \lambda \right) }$ and $f_{D/2-1-l}$ are small. By using the
corresponding asymptotic expressions one can show that the dominant
contribution comes from the term $m=0$ and $\lambda =0$. For even values of $%
D>4$, to the leading order one gets the following expressions (no summation
over $i$)%
\begin{eqnarray}
\langle T_{i}^{i}\rangle _{\mathrm{b}} &\approx &\frac{2^{D-2}\Gamma
^{2}(D/2+1)\Gamma ^{2}(D/2)}{(D-2)\left( D-4\right) \Gamma (D+2)}\frac{%
\alpha B_{D}C_{D}^{(i)}}{(\alpha r/\eta )^{D+2}\ln (r/a)},  \notag \\
\langle T_{0}^{1}\rangle _{\mathrm{b}} &\approx &\frac{2^{D-4}D\Gamma
^{4}(D/2)B_{D}}{\Gamma (D)(\alpha r/\eta )^{D+1}\ln (r/a)},  \label{Tiilarge}
\end{eqnarray}%
with the coefficients%
\begin{eqnarray}
C_{D}^{(0)} &=&D(D-1)\left( 6-D\right) -6,  \notag \\
C_{D}^{(1)} &=&D\left( D-6\right) +2,  \notag \\
C_{D}^{(2)} &=&D^{2}\left( 3-D\right) -2,  \notag \\
C_{D}^{(l)} &=&D^{2}\left( 6-D\right) -8D-6,  \label{CiD}
\end{eqnarray}%
$l=3,\ldots ,D$. The corresponding energy density is positive. by taking
into account that near the shell the energy density is negative for $%
D\geqslant 4$, we conclude that at some intermediate value of the radial
coordinate the boundary-induced contribution in the energy density vanishes.
The asymptotic (\ref{Tiilarge}) for the energy flux is valid in the case $%
D=4 $ as well. For $D=4$, the leading term in the expansion of the diagonal
components is given by (no summation over $i$)%
\begin{equation}
\langle T_{i}^{i}\rangle _{\mathrm{b}}\approx \frac{\alpha \ln (r/\eta
)C_{4}^{(i)}}{20\pi ^{3}(\alpha r/\eta )^{6}\ln (r/a)},  \label{TiilargeD4}
\end{equation}%
where $C_{4}^{(0)}=-C_{4}^{(2)}=3$, and $C_{4}^{(i)}=-1$ for$\;i=1,3,4$. For
the Minkowski bulk, the large distance asymptotic is given by (no summation
over $i$)
\begin{equation}
\langle T_{i}^{i}\rangle _{\mathrm{b}}^{(\mathrm{M})}\approx \frac{\pi
^{-(D+1)/2}\left( 2-D\right) \Gamma ^{3}((D+1)/2)}{4r^{D+1}\ln (r/a)\left(
D-1\right) \Gamma (D+1)}C_{(\mathrm{M})}^{(i)},  \label{TiiMlarge}
\end{equation}%
where%
\begin{eqnarray}
C_{(\mathrm{M})}^{(0)} &=&D^{2}-4D+1,\;i=0,3,\ldots ,D,  \notag \\
C_{(\mathrm{M})}^{(1)} &=&1-D,\;C_{(\mathrm{M})}^{(2)}=D(D-1).  \label{CiM}
\end{eqnarray}%
The corresponding energy density is positive for $D=3$ and negative for $%
D\geqslant 4$.

The shell-induced contribution to the energy density and the energy flux in
the exterior region are plotted in figure \ref{fig2} for the $D=4$ model.
The large distance asymptotic is given by (\ref{TiilargeD4}) and the energy
density is positive. For points near the shell we have the asymptotic
behavior (\ref{Tiinear}), with the replacement $a-r\rightarrow r-a$, and the
energy density is negative. Note that for the $D=4$ Minkowski bulk the
corresponding energy density is negative everywhere. For the $D=3$ model,
the energy density is positive in the exterior region and negative in the
interior region.

In the discussion above we have considered the boundary condition (\ref{BC1}%
) that generalizes the condition at the surface of a conductor for arbitrary
number of spatial dimensions. Another type of boundary conditions for a
gauge field is used in bag models of hadrons and in flux tube models of
confinement in quantum chromodynamics (see, for instance, \cite{Cand86}).
This boundary condition has the form%
\begin{equation}
n^{\mu }F_{\mu \nu }=0,  \label{BC2}
\end{equation}%
on the boundary of a volume inside of which the gluons are confined. In flux
tube models the gauge field is confined inside a cylinder. The corresponding
Casimir densities inside and outside a cylindrical shell are investigated in
a way similar to that we have described above. The mode functions still have
the form (\ref{A1}) and (\ref{A2}). Imposing the boundary condition (\ref%
{BC2}), we can see that, in the interior region, eigenvalues for $\gamma $
are roots of the equation (\ref{BCb}) for the mode $\sigma =1$ and the roots
of (\ref{BCa}) for $\sigma =2,\ldots ,D-1$. The final expressions for the
VEVs are obtained from those given above by the replacement%
\begin{equation}
\dfrac{K_{m}^{\left( \lambda \right) }\left( xa/\eta \right) }{I_{m}^{\left(
\lambda \right) }\left( xa/\eta \right) }\rightarrow \dfrac{K_{m}^{\left(
1-\lambda \right) }\left( xa/\eta \right) }{I_{m}^{\left( 1-\lambda \right)
}\left( xa/\eta \right) },  \label{ReplQCD}
\end{equation}%
in both the interior and exterior regions. In particular, the VEV of the
squared electric field is negative in these regions.

\section{Conclusion}

\label{sec:Conc}

In the investigations of the Casimir effect the cylindrically symmetric
boundaries are among the most popular geometries. In the present paper we
have investigated the local Casimir densities for the electromagnetic field
inside and outside a cylindrical shell in background of $(D+1)$-dimensional
dS spacetime. On the shell, the field tensor obeys the boundary condition (%
\ref{BC1}). In the special case $D=3$ this corresponds to the perfect
conductor boundary condition. The procedure, we employed for the evaluation
of the VEVs bilinear in the field, is based on the mode-sum formula (\ref%
{VEV}). In this procedure the complete set of cylindrical mode functions for
the electromagnetic field, obeying the boundary condition, is required. In
the problem under consideration one has a single mode of the TE type and $%
D-2 $ modes of the TM type. For the Bunch-Davies vacuum state the
corresponding vector potentials are given by the expressions (\ref{A1}) and (%
\ref{A2}) with the radial functions (\ref{RadIn}) and (\ref{RadOut}) for the
exterior and interior regions, respectively.

We have investigated the combined effects of a cylindrical boundary and
background gravitational field on the VEVs of the electric field squared and
of the energy-momentum tensor. In the interior region the eigenvalues of the
quantum number $\gamma $ are expressed in terms of the zeros of the Bessel
function $J_{m}(x)$ for the TM modes and in terms of the zeros of the
derivative $J_{m}^{\prime }(x)$ in the case of the TE mode. For the
summation of the series over these zeros we have used the generalized
Abel-Plana summation formula (\ref{SumForm}). This allowed us to extract
from the VEVs the parts corresponding to the boundary-free dS spacetime and
to present the shell-induced contributions in terms of strongly convergent
integrals, for points away from the boundary. With this separation, the
renormalization of the VEVs is reduced to the one for the boundary-free
geometry. As a result, inside the shell, the VEVs are decomposed as (\ref%
{E2Dec}) and (\ref{Tdec}) with the shell-induced parts given by (\ref{E2b})
for the electric field squared and by (\ref{Tiib}) for the diagonal
components of the energy-momentum tensor. A similar decomposition is
provided for the exterior region. The expressions for the shell-induced
parts in this region differ from the ones inside the shell by the
replacements $I_{m}\rightleftarrows K_{m}$ of the modified Bessel functions
(see (\ref{E2bout2}) and (\ref{Tiibout})). For both the interior and
exterior regions the shell-induced contributions to the VEV of the electric
field squared are positive. In addition to the diagonal components, the VEV
of the energy-momentum tensor has nonzero off-diagonal component $\langle
T_{0}^{1}\rangle $. It corresponds to the energy flux along the radial
direction and is given by the expressions (\ref{T01b}) and (\ref{T10bout})
for the exterior and interior regions. The off-diagonal component is
negative inside the shell and positive in the exterior region. This means
that the energy flux is directed from the shell in both the regions. On the
axis of the shell the flux vanishes.

We have considered various special cases of general formulas. In the limit $%
\alpha \rightarrow \infty $, the VEVs inside and outside a cylindrical shell
in the background of $(D+1)$-dimensional Minkowski spacetime are obtained.
The corresponding expressions generalize the results previously known for $%
D=3$ to an arbitrary number of spatial dimensions. Note that for $D=3$ the
electromagnetic field is conformally invariant and the shell-induced VEVs in
the dS bulk are obtained from those in Minkowski spacetime by the standard
conformal transformation. For points near the cylindrical boundary the
contribution of small wavelengths dominates in the shell-induced VEVs. The
leading terms in the corresponding asymptotic expansions for the field
squared and diagonal components of the energy-momentum tensor coincide with
those for a cylindrical shell in the Minkowski bulk with the distance from
the shell replaced by the proper distance in dS bulk. The leading term in
the energy flux is given by the relation (\ref{T10near}). The effects of the
background gravitational field on the shell-induced VEVs are essential at
distances from the boundary larger than the curvature radius of the dS
spacetime. In particular, for the numerical example considered by us in the
case $D=4$, at large distances the shell-induced contribution to the vacuum
energy density is negative for the Minkowski bulk and positive for dS
background. Near the shell, the energy density is negative in both these
cases. As a consequence, for the dS bulk it vanishes for some intermediate
value of the radial coordinate.

Another boundary condition, used for the confinement of gauge fields in bag
models of hadrons and in flux tube models of QCD, is the one given by (\ref%
{BC2}). The corresponding expressions for the VEVs of the field squared and
energy-momentum tensor are obtained from those for generalized perfect
conductor boundary condition by the replacement (\ref{ReplQCD}). In this
case, the boundary-induced contribution on the VEV of the squared electric
field is neagtive in both the interior and exterior regions.

\section*{Acknowledgments}

A. A. S. and N. A. S. were supported by the State Committee of Science
Ministry of Education and Science RA, within the frame of Grant No. SCS
15T-1C110, and by the Armenian National Science and Education Fund (ANSEF)
Grant No. hepth-4172.

\appendix

\section{Cylindrical modes in Minkowski spacetime}

\label{sec:App1}

In this section we consider the cylindrical modes for the electromagnetic
field in $(D+1)$-dimensional Minkowski spacetime. For the electromagnetic
field one has $D-1$ polarization states specified by $\sigma =1,2,\ldots
,D-1 $. The vector potential for the polarization $\sigma $ will be denoted
by $A_{\sigma \mu }$, $\mu =0,1,\ldots ,D$. We will impose the gauge
condition $\nabla _{(\mathrm{M})\mu }A_{\sigma }^{\mu }=0$, where $\nabla _{(%
\mathrm{M})\mu }$ is the covariant derivative operator associated with the
Minkowskian metric tensor $g_{(\mathrm{M})\mu \nu }=\mathrm{diag}%
(1,-1,-r^{2},-1,\ldots ,-1)$. From the field equation $\nabla _{(\mathrm{M}%
)\mu }F^{\mu \nu }=0$ one gets%
\begin{equation}
\left( \Delta -\partial _{0}^{2}\right) A_{\sigma \mu }=0,  \label{AeqM}
\end{equation}%
for $\mu =0,3,\ldots ,D$, and%
\begin{eqnarray}
\left( \partial _{0}^{2}-\Delta \right) A_{\sigma 1}+\frac{A_{\sigma 1}}{%
r^{2}}+\frac{2}{r^{3}}\partial _{2}A_{\sigma 2} &=&0,  \notag \\
\partial _{0}^{2}A_{\sigma 2}-\Delta A_{\sigma 2}+\frac{2}{r}\partial
_{1}A_{\sigma 2}-\frac{2}{r}\partial _{2}A_{\sigma 1} &=&0,  \label{AeqM2}
\end{eqnarray}%
for the radial and azimuthal components. Here%
\begin{equation}
\Delta =\frac{1}{r}\partial _{1}\left( r\partial _{1}\right) +\frac{1}{r^{2}}%
\partial _{2}^{2}+\sum_{l=3}^{D}\partial _{l}^{2}.  \label{Lap}
\end{equation}

For the polarization $\sigma =1$ we take
\begin{equation}
A_{\sigma \mu }=(0,-r^{-1}\partial _{2},r\partial _{1},0,\ldots ,0)\psi
_{\sigma },\;\sigma =1.  \label{A1M}
\end{equation}%
It can be easily checked that this function obeys the gauge condition. The
field equations (\ref{AeqM2}) are satisfied if the function $\psi _{\sigma }$
obeys the equation%
\begin{equation}
\left( \Delta -\partial _{0}^{2}\right) \psi _{\sigma }=0.  \label{psieqM}
\end{equation}

For the polarizations $\sigma =2,\ldots ,D-1$ we present the vector
potential in the form%
\begin{equation}
A_{\sigma \mu }=\epsilon _{\sigma \mu }\psi _{\sigma },\;\sigma =2,\ldots
,D-1,  \label{A2M}
\end{equation}%
with scalar functions $\psi _{\sigma }$ and $\epsilon _{\sigma 1}=\epsilon
_{\sigma 2}=0$. From the field equations (\ref{AeqM}) it follows that the
functions $\psi _{\sigma }$ obey the wave equation (\ref{psieqM}). The gauge
condition is reduced to%
\begin{equation}
g_{(\mathrm{M})}^{\mu \nu }\epsilon _{\sigma \mu }\partial _{\nu }\psi
_{\sigma }=0.  \label{Gauge2}
\end{equation}

The solutions for the scalar functions $\psi _{\sigma }$, $\sigma
=1,2,\ldots ,D-1$, have the form%
\begin{equation}
\psi _{\sigma }=C_{m}(\gamma r)e^{i(m\phi +\mathbf{k}\cdot \mathbf{z}-\omega
t)},  \label{psi}
\end{equation}%
where $C_{m}(x)$ is a cylinder function of the order $m=0,\pm 1,\pm 2,\ldots
$, $\mathbf{k}\cdot \mathbf{z}=\sum_{l=3}^{D}k_{l}z^{l}$, and $\omega $ is
given by (\ref{om}). We will normalize the polarization vectors in
accordance with the relation%
\begin{equation}
g_{(\mathrm{M})}^{\mu \nu }\epsilon _{\sigma \mu }\epsilon _{\sigma ^{\prime
}\nu }=-\frac{\gamma ^{2}}{\omega ^{2}}\delta _{\sigma \sigma ^{\prime }}.
\label{epsnorm}
\end{equation}%
From the gauge condition one has $\epsilon _{\sigma 0}=-\mathbf{k}\cdot
\mathbf{\epsilon }_{\sigma }/\omega $, where $\mathbf{k}\cdot \mathbf{%
\epsilon }_{\sigma }=\sum_{l=3}^{D}k_{l}\epsilon _{\sigma l}$. Combining
with (\ref{epsnorm}) the following relations are obtained:%
\begin{equation}
\sum_{l,n=3}^{D}\left( \omega ^{2}\delta _{nl}-k_{l}k_{n}\right) \epsilon
_{\sigma l}\epsilon _{\sigma ^{\prime }n}=\gamma ^{2}\delta _{\sigma \sigma
^{\prime }},  \label{rel1M}
\end{equation}%
and%
\begin{equation}
\sum_{\sigma =2}^{D-1}\epsilon _{\sigma l}\epsilon _{\sigma n}=\omega
^{-2}(k_{l}k_{n}+\gamma ^{2}\delta _{ln}),  \label{rel2M}
\end{equation}%
for $l,n=3,\ldots ,D$.

An alternative form for the cylindrical modes (\ref{A2M}) with the
polarizations $\sigma =2,\ldots ,D-1$ is obtained by making the gauge
transformation%
\begin{equation}
A_{\sigma \mu }^{\prime }=A_{\sigma \mu }+\partial _{\mu }f_{\sigma },
\label{AnewM}
\end{equation}%
with the function%
\begin{equation}
f_{\sigma }=i\omega ^{-2}\mathbf{k}\cdot \mathbf{\epsilon }_{\sigma }\psi
_{\sigma }.  \label{fgauge}
\end{equation}%
In the new gauge the scalar potential vanishes, $A_{\sigma 0}^{\prime }=0$,
and one has%
\begin{equation}
A_{\sigma \mu }^{\prime }=\left( 0,\left( \epsilon _{\sigma l}+i\omega ^{-2}%
\mathbf{k}\cdot \mathbf{\epsilon }_{\sigma }\partial _{l}\right) \psi
_{\sigma }\right) ,\;l=1,\ldots ,D.  \label{AmuM}
\end{equation}

Hence, for $(D+1)$-dimensional Minkowski spacetime the cylindrical modes for
the electromagnetic field in the gauge $A_{\sigma 0}=0$, $\partial
_{l}(rA^{l})=0$ are given by (\ref{A1M}) and (\ref{AmuM}), where for the
scalar function $\psi _{\sigma }$ one has the expression (\ref{psi}). The
radial function $C_{m}(\gamma r)$ is a linear combination of the Bessel and
Neumann functions. The relative coefficient in this linear combination
depends on the specific problem. For example, inside a cylindrical shell one
has $C_{m}(\gamma r)\sim J_{m}(\gamma r)$. In the special case $D=3$, the
modes (\ref{A1M}) and (\ref{AmuM}) are reduced to the well known TE and TM
modes in cylindrical waveguides (see, for instance, \cite{Jack99}). In the $%
(D+1)$-dimensional case we have a single mode of the TE type and $D-2$ modes
of the TM\ type. The corresponding boundary conditions are discussed in
section \ref{sec:Modes}.

\section{Evaluation of the integrals in the model $D=4$}

\label{sec:App2}

In this section we evaluate the integrals of the form (\ref{Ical}) appearing
in the expressions for the VEVs in the special case $D=4$. First of all, by
using the integration formula from \cite{Prud86} we can see that%
\begin{eqnarray}
\mathcal{I}_{1,0}(x) &=&\frac{1}{2}\left[ f_{0}(x)+f_{1}(x)\right] ,  \notag
\\
\mathcal{I}_{1,1}(x) &=&\mathcal{I}_{1,0}(x)-\frac{2}{x}I_{1}(x)K_{0}(x).
\label{Int0}
\end{eqnarray}%
Our starting point in the evaluation of the remaining integrals is the
formula (see, for instance, \cite{Prud86})%
\begin{equation}
\int_{0}^{1}ds\,sI_{0}\left( xs\right) K_{0}\left( ys\right) =-\frac{%
xI_{1}\left( x\right) K_{0}\left( y\right) +bI_{0}\left( x\right)
K_{1}\left( y\right) -1}{y^{2}-x^{2}}.  \label{Int1}
\end{equation}%
Applying the operator $-\lim_{y\rightarrow x}\partial _{y}$ on the left and
right hand sides of this formula we find%
\begin{equation}
\int_{0}^{1}ds\,s^{2}I_{0}\left( xs\right) K_{1}\left( xs\right) =\frac{%
f_{1}\left( x\right) +1}{4x}.  \label{Int11}
\end{equation}%
Next, combining with the relation $I_{0}\left( x\right) K_{1}\left( x\right)
=1/x-I_{1}\left( x\right) K_{0}\left( x\right) $ one gets%
\begin{equation}
\int_{0}^{1}dss^{2}I_{1}\left( xs\right) K_{0}\left( xs\right) =\frac{%
1-f_{1}\left( x\right) }{4x}.\,  \label{Int12}
\end{equation}

For the evaluation of the integral $\mathcal{I}_{3,1}(x)$ we apply the
operator $-2\lim_{y\rightarrow x}\partial _{x}\partial _{y}$ on the left-
and right-hand sides of (\ref{Int1}). This gives%
\begin{equation}
\mathcal{I}_{3,1}(x)=\frac{2}{3x^{2}}\left[ 1-2xI_{1}\left( x\right)
K_{0}\left( x\right) \right] +\frac{1}{6}\left[ f_{0}\left( x\right) +\left(
1-4/x^{2}\right) f_{1}\left( x\right) \right] .  \label{Int2}
\end{equation}%
Integrating by parts the integral $\int_{0}^{1}ds\,s^{3}I_{0}^{\prime
}\left( xs\right) K_{0}^{\prime }\left( xs\right) $ we can see that%
\begin{equation}
\mathcal{I}_{3,0}(x)=\mathcal{I}_{3,1}(x)-\frac{2}{x}I_{0}\left( x\right)
\,K_{1}\left( x\right) +\frac{4}{x}\int_{0}^{1}dss^{2}I_{0}\left( sx\right)
K_{1}\left( sx\right) .  \label{Int3}
\end{equation}%
By taking into account the relations (\ref{Int11}) and (\ref{Int2}) we obtain%
\begin{equation}
\mathcal{I}_{3,0}(x)=\frac{1}{3x^{2}}\left[ 2xI_{1}\left( x\right)
K_{0}\left( x\right) -1\right] +\frac{1}{6}\left[ f_{0}\left( x\right)
+\left( 1+2/x^{2}\right) f_{1}\left( x\right) \right] \,.  \label{Int4}
\end{equation}

\end{document}